\documentclass[11pt,a4paper]{article}
\pdfoutput=1
\usepackage{jheppub}

\def\p{{\mathbf p}}
\def\s{{\mathbf s}}
\def\q{{\mathbf q}}
\def\h{{\mathbf h}}
\def\k{{\mathbf k}}
\def\x{{\mathbf x}}
\def\y{{\mathbf y}}
\def\yb{\bar{\mathbf y}}
\def\r{{\mathbf r}}
\def\z{{\mathbf z}}
\def\zb{\bar{\mathbf z}}
\def\v{{\mathbf v}}
\def\u{{\mathbf u}}
\def\b{{\mathbf b}}
\def\B{{\mathbf B}}
\def\G{{\cal G}}
\def\U{\mathcal{U}}
\def\P{{\cal P}}

\newcommand{\beq}{\begin{eqnarray}}
\newcommand{\eeq}{\end{eqnarray}}
\newcommand{\be}{\begin{eqnarray*}}
\newcommand{\ee}{\end{eqnarray*}}

\DeclareMathOperator{\tr}{tr}

\title{Next-to-eikonal corrections in the CGC: \hspace{3cm}
  gluon production and spin asymmetries in pA collisions}
\author[a]{Tolga Altinoluk,}
\author[a]{N\'estor Armesto,}
\author[a]{Guillaume Beuf,}
\author[b]{Mauricio Mart\'inez}
\author[a]{and Carlos A. Salgado}
\affiliation[a]{Departamento de F\'isica de Part\'iculas and IGFAE,
Universidade de Santiago de Compostela,
E-15706 Santiago de Compostela,
Galicia-Spain}
\affiliation[b]{Department of Physics, The Ohio State University, Columbus, OH 43210, USA}

\emailAdd{tolga.altinoluk@usc.es}
\emailAdd{nestor.armesto@usc.es}
\emailAdd{guillaume.beuf@usc.es}
\emailAdd{martinezguerrero.1@osu.edu}
\emailAdd{carlos.salgado@usc.es}

\abstract{We present a new method to systematically include corrections to the eikonal approximation in the background field formalism. Specifically, we calculate the subleading, power-suppressed corrections due to the finite width of the target or the finite energy of the projectile.
Such power-suppressed corrections involve Wilson lines decorated by gradients of the background field - thus related to the density - of the target. The method is of generic applicability. As a first example, we study single inclusive gluon production in pA collisions, and various related spin asymmetries, beyond the eikonal accuracy.
}

\begin{document}
\maketitle

\pagestyle{empty}
\newpage

\mbox{}

\pagestyle{plain}

\setcounter{page}{1}

\section{Introduction}
\label{sec:intro}

High-energy scattering is usually treated in the eikonal approximation.
Generically,  each of the constituents of the projectile picks up a phase (in QED \cite{Bjorken:1970ah} or in gravity \cite{'tHooft:1987rb} for example) or a Wilson line (in Yang-Mills theories \cite{Collins:1985gm}) upon scattering on a background field representing the target. The phases or the Wilson lines probe this field at a given transverse point but integrated along the light-cone
direction of propagation of the projectile. This is a consequence of the large boost performed on the target, which appears to the projectile as infinitely Lorentz contracted.

In the case of
the scattering of a dilute projectile on a dense target, the high density of the target makes it possible to perform a semi-classical approximation that amounts to replace the target by an intense and random classical background field. 
The Color Glass Condensate (CGC, see \cite{Gelis:2010nm} and references therein) is the effective theory that allows one to study such scatterings in high-energy Yang-Mills theories, within the eikonal and semi-classical approximations. In the CGC,  one can calculate observables for such processes in a weak coupling expansion, and resum high-energy leading and next-to-leading logarithms. Corrections suppressed by inverse powers of the energy of the collision are systematically neglected within the eikonal limit. This situation can be paralleled in the case of the perturbative QCD description of hard processes. There,  one can calculate systematically perturbative corrections to the relevant observables, and resum leading collinear logarithms within the collinear factorization framework, whereas the corrections suppressed by inverse powers of the hard scale are usually neglected. However, these power-suppressed terms  can be sizeable in practical applications. The leading ones have been studied explicitly for various observables, see e.g. \cite{Efremov:1984ip,Qiu:1991pp,Qiu:1998ia,Bacchetta:2006tn,Brodsky:2002cx,Barone:2010zz,Boer:2011fh} and references therein. Noticeably, some of these terms are negligible except for less inclusive situations such as semi-inclusive production or polarized collisions.

The aim of the present paper is to study one type of such power-suppressed corrections to the CGC, namely the next-to-eikonal contributions\footnote{ Next-to-eikonal corrections have been studied in the context of soft-gluon resummation for scattering amplitudes in Ref.\cite{Laenen:2008gt}. However, the connection between that study and ours is not clear due to their different scope and the different formalisms employed.} due to the finite length of the target or the large but finite energy of the projectile. They can be understood as the subleading effects with respect to either infinite Lorentz contraction of the target or infinite Lorentz dilation of the projectile.
For the production of a gluon with momentum $k=(k^+,k^-,\k)$ on a target of light-cone thickness $L^+$, the eikonal expansion amounts to assume that $\sqrt{k^+/L^+}$ is larger than any available transverse momentum scale, like the transverse momentum of the produced gluon, $\k$,  or the saturation scale of the target. Next-to-eikonal corrections are then suppressed as $L^+/k^+$ compared to the strict eikonal terms.

As an application, we focus on particle production in nuclear collisions at high energies. This is a very interesting problem both from a theoretical point of view as it provides a testing ground for our understanding of QCD and its factorization properties, and from a phenomenological point of view due to its application to experimental studies of such collisions. While the existence of a truly hard scale demands extensions of the usual collinear factorization theorems (see \cite{Collins:1989gx} and references therein) to the nuclear case, the situation for a semi-hard scale is far less clear (the so-called $k_\perp$-factorization \cite{Catani:1990eg,Levin:1991ry}). On the other hand, it is the latter that is expected to determine the bulk of particle production in high-energy collisions. It should also provide the initial conditions for the evolution of the medium produced in ultra relativistic nuclear collisions.
Several calculations exist for computing the single inclusive \cite{Kovchegov:1998bi,Kovchegov:2001sc,Dumitru:2001ux,Kharzeev:2003wz,Blaizot:2004wu} gluon
cross section in the semi-hard regime in pA collisions at high energy, resulting in a $k_\perp$-factorized expression.

Let us anticipate the results and conclusions of our study. 
In the calculation of dense-dilute scattering processes in the high-energy limit of QCD, the main building blocks sensitive to the finite length of the target are the propagators of gluons and quarks in the background field~\cite{Wiedemann:2000za,Zakharov:1996fv,Baier:1996kr,Zakharov:1998sv,MehtarTani:2006xq}, used previously in studies of jet quenching. 
As a first step, a systematic eikonal expansion of the retarded gluon propagator in a background field is derived. The leading term of the expansion, which corresponds to the strict eikonal limit, naturally involves the usual Wilson line operator. However, the next-to-eikonal terms of the expansion give rise to new type of operators that contain gradients of the usual Wilson line operators. The explicit form of these new operators is written as
\beq
\label{decoratedWilson1}
\U^{i,ab}_{(1)}(x^+,y^+,\y)=\int_{y^+}^{x^+}dz^+\frac{1}{(x^+-y^+)}\bigg\{\bigl[\partial_{\y^i}\U(x^+,z^+,\y)\bigr]
\U(z^+,y^+,\y)\bigg\}^{ab}
\eeq
and
\beq
\label{decoratedWilson2}
\U^{ab}_{(2)}(x^+,y^+,\y)=\int_{y^+}^{x^+}dz^+\frac{1}{(x^+-y^+)}\bigg\{\bigl[\partial^2_{\y}\U(x^+,z^+,\y)\bigr]
\U(z^+,y^+,\y)\bigg\}^{ab}.
\eeq


This expansion can be used for different high-energy processes like DIS or single inclusive particle production in both the $k_\perp$-factorization    \cite{Kovchegov:1998bi,Kovchegov:2001sc,Dumitru:2001ux,Kharzeev:2003wz,Blaizot:2004wu} and the hybrid formalism \cite{Dumitru:2005gt,Altinoluk:2011qy,Chirilli:2012jd,Kang:2014lha}, and for the calculation of medium-induced gluon radiation. In this paper, we concentrate on single inclusive gluon production in pA collisions at central rapidity, far from the target and projectile fragmentation regions. 
In unpolarized pA scattering, the next-to-eikonal correction is found 
to be identically zero, thus extending the validity of the usual $k_\perp$-factorized formula. 
On the other hand, for various types of spin asymmetries, the eikonal contribution vanishes, leaving the next-to-eikonal correction as the leading term. This is the case for instance for single transverse spin asymmetry with a polarized target, or for polarized gluon (or hadron) production in unpolarized pA collisions.
In the case of the single inclusive gluon spectrum for pA collisions at a given impact parameter, both the eikonal and the next-to-eikonal contributions can be non-vanishing, depending on the adopted formalism and approximations. 

The plan of the paper is as follows: we perform the eikonal expansion of the gluon propagator in the background field and obtain the first corrections with respect to the high energy limit in Sec. \ref{sec:G_expansion}.  Then, we apply this result to the case of single inclusive gluon production at central rapidities in pA collisions. In Sec. \ref{sec:kt_fact_beyond_eikonal}, we calculate this observable for a target of finite size as a function of the gluon propagator in the background field and, in Sec. \ref{sec:kt_fact_next2eikonal}, we use the expansion of the propagator obtained in Sec. \ref{sec:G_expansion} in order to derive next-to-eikonal corrections to gluon production observables including various spin asymmetries. Finally, we present our conclusions in Sec. \ref{sec:conclusions}.

\section{Eikonal expansion of the retarded gluon propagator in a background field}
\label{sec:G_expansion}

One of the main building blocks for dense-dilute scattering processes at high energy is the retarded gluon propagator in a classical background field representing a fast nucleus. In this section, we recall the general properties of that object in light-cone gauge, and then present a method to calculate it systematically in an eikonal expansion, that we evaluate explicitly to first order.

\subsection{Properties of the background gluon propagator}
A highly boosted left-moving target can be described by a classical gluon background field ${\cal A}_a^{-}(x^+,\x)$. Only the $(-)$ component of this field is enhanced by a Lorentz gamma factor, so that the other components are negligible in comparison. Moreover, due to time dilation, the $x^-$ dependence of the field can be neglected.


In the presence of such a background field, it is natural to work in the light-cone gauge $A^+_a=0$.
In this gauge, the tensor structure of the free gluon propagators guaranties that
\begin{equation}
G^{+\nu}_{0}(x,y)=G^{\mu+}_{0}(x,y)=0\, .
\end{equation}
The background propagators have to start and finish with a free propagator, which imposes that 
\begin{equation}
G^{+\nu}(x,y)=G^{\mu+}(x,y)=0\, .
\end{equation}
Linearizing the Yang-Mills equations around the background field ${\cal A}_a^{-}(x^+,\x)$, one finds the equation satisfied by the gluon propagators in that background
\begin{eqnarray}
& &\!\!\!\!\!\!\!\!\! \bigg[g_{\mu \nu} \left(\delta_{ab}\, \Box_{x}-2ig \Big({\cal A}^{-}(x^+,\x)\cdot T\Big)_{\! ab}\partial_{x^-}  \right)- \delta_{ab} \partial_{x^\mu} \partial_{x^\nu} \bigg] G^{\nu\rho}(x,y)_{bc}\nonumber\\
& &=i\: {g_{\mu}}^{\rho}\, \delta_{ac}\: \delta^{(4)}(x\!-\!y)   \, , \qquad \textrm{for } \; \mu\neq +\, .\label{GreensEqVector}
\end{eqnarray}

For our purposes, we need only the retarded propagator. The retarded solution of the Green's equation, Eq. \eqref{GreensEqVector}, written as a function of the retarded scalar propagator $G_{R}(x,y)_{ab}$, is given as
\begin{eqnarray}
G^{ij}_{R}(x,y)_{ab}&=&\delta^{ij}\, G_{R}(x,y)_{ab}\ ,\label{VectPerpPerp2ScalPos}\\
G^{-i}_{R}(x,y)_{ab}&=&-\partial_{\x^i}\, \int_{-\infty}^{x^-}\!\!\! \textrm{d}z^-\; G_{R}(x^+,z^-,\x;y)_{ab}\ ,\label{VectMinPerp2ScalPos}\\
G^{i-}_{R}(x,y)_{ab}&=&\partial_{\y^i}\, \int_{y^-}^{+\infty}\!\!\! \textrm{d}z^-\; G_{R}(x;y^+,z^-,\y)_{ab}\ ,\label{VectPerpMin2ScalPos}\\
G^{--}_{R}(x,y)_{ab}&=&-\partial_{\x^i}\partial_{\y^i}\, \int_{-\infty}^{x^-}\!\!\! \textrm{d}w^-\; \int_{y^-}^{+\infty}\!\!\! \textrm{d}z^-\; G_{R}(x^+,w^-,\x;y^+,z^-,\y)_{ab}\nonumber\\
& &-i\:\delta_{ab}\; (x^-\!-\!y^-)\, \theta(x^-\!-\!y^-)\, \delta(x^+\!-\!y^+)\, \delta^{(2)}(\x\!-\!\y)\,   ,\label{VectMinMin2ScalPos}
\end{eqnarray}
where $G_{R}(x,y)_{ab}$ is the retarded solution of the equation
\begin{eqnarray}
\left(\delta_{ab} \, \Box_{x}-2ig \Big({\cal A}^{-}(x^+,\x)\cdot T\Big)_{\! ab}\partial_{x^-}  \right) G_{R}(x,y)_{bc}&=&-i\:\delta_{ac}\: \delta^{(4)}(x\!-\!y)\, . \label{GreensEqScal}
\end{eqnarray}
Since the background field is independent of $x^-$, it is convenient to introduce the Fourier transform of the gluon and scalar propagators\footnote{Here and after, we use the notation $\underline{x}=(x^+,\x)$.}
\begin{equation}
G^{(\mu\nu)}_{R}(x,y)_{ab}=\int \frac{\textrm{d}p^+}{2\pi}\, e^{-ip^+ (x^-\!-\!y^-)}\; \frac{1}{2 (p^+\!+\!i\epsilon)}  \,{\cal G}^{(\mu\nu)}_{p^+}(\underline{x};\underline{y})_{ab}\, ,\label{FTPropMixed}
\end{equation}
allowing to rewrite the relations given in Eqs. \eqref{VectPerpPerp2ScalPos}, \eqref{VectMinPerp2ScalPos}, \eqref{VectPerpMin2ScalPos} and \eqref{VectMinMin2ScalPos} as
\begin{eqnarray}
{\cal G}^{ij}_{k^+}(\underline{x};\underline{y})^{ab}&=&\delta^{ij}\, {\cal G}^{ab}_{k^+}(\underline{x};\underline{y}),\label{VectPerpPerp2ScalMix}\\
{\cal G}^{-i}_{k^+}(\underline{x};\underline{y})^{ab}&=&\frac{-i}{k^+\!+\!i\epsilon}\;  \partial_{\x^i}\,  {\cal G}^{ab}_{k^+}(\underline{x};\underline{y}),\label{VectMinPerp2ScalMix}\\
{\cal G}^{i-}_{k^+}(\underline{x};\underline{y})^{ab}&=&\frac{i}{k^+\!+\!i\epsilon}\;  \partial_{\y^i}\,  {\cal G}^{ab}_{k^+}(\underline{x};\underline{y}),\label{VectPerpMin2ScalMix}\\
{\cal G}^{--}_{k^+}(\underline{x};\underline{y})^{ab}&=&\!\!\frac{1}{(k^+\!+\!i\epsilon)^2}\;  \partial_{\x^i}\,\partial_{\y^i}\,  {\cal G}^{ab}_{k^+}(\underline{x};\underline{y})\!+\!\frac{2i}{k^+\!+\!i\epsilon}\, \delta^{ab}\;  \delta^{(3)}\!(\underline{x}\!-\!\underline{y})  \label{VectMinMin2ScalMix}
\end{eqnarray}
and the scalar Green's equation, Eq. \eqref{GreensEqScal}, as
\begin{equation}
\bigg[\delta^{ab} \,\left(i\partial_{x^+}+\frac{\partial_{\x}^2}{2(k^+\!+\!i\epsilon)} \right)   +g \Big({\cal A}^{-}(\underline{x})\cdot T\Big)^{\! ab} \bigg] {\cal G}^{bc}_{k^+}(\underline{x};\underline{y})= i\, \delta^{ac}\: \delta^{(3)}(\underline{x}\!-\!\underline{y})\, . \label{SchroEq}
\end{equation}
This is the Green's equation associated with a Schr\"odinger equation in $2+1$ dimensions, for a particle of \emph{mass} $k^+\!+\!i\epsilon$ in a space-time dependent matrix potential $-g \Big({\cal A}^{-}(\underline{x})\cdot T\Big)^{\! ab}$, $T\cdot A\equiv T^a A_a$ in the adjoint representation.
The solution of Eq.\eqref{SchroEq} can be written formally as a path integral~\cite{Wiedemann:2000za,Zakharov:1996fv,Baier:1996kr,Zakharov:1998sv,MehtarTani:2006xq}
\begin{eqnarray}
&&\hspace{-1.6cm}{\cal G}^{ab}_{k^+}(\underline{x};\underline{y})=   \theta(x^+\!-\!y^+)\, \int_{\z(y^+)=\y}^{\z(x^+)=\x} {\cal D}\z(z^+)\;  \exp \Bigg[ \frac{i k^+}{2} \int_{y^+}^{x^+} dz^+\;  \dot{\z}^2(z^+) \Bigg]\;\nonumber\\
&& \hspace{2cm} \times \:\:\:\U^{ab}\Big(x^+,y^+,\big[\z(z^+)\big]\Big)
\; ,  \label{PathIntDef_cont}
\end{eqnarray}
with the Wilson line
\begin{eqnarray}
\U^{ab}\Big(x^+,y^+,\big[\z(z^+)\big]\Big)&=& {\cal P}_{+}  \exp  \Bigg\{ ig \int_{y^+}^{x^+} dz^+\;  T\cdot {\cal A}^{-} \Big(z^+,\z(z^+) \Big)\Bigg\}^{ab}
 \label{BrownWilsonLine}
\end{eqnarray}
following the Brownian trajectory $\z(z^+)$. In the last expression ${\cal P}_{+}$ indicates the ordering of color generators $T^a$ along $x^+$.
The path integral, Eq.\eqref{PathIntDef_cont}, is actually defined through discretization as
\begin{eqnarray}
&&\hspace{-1.6cm}{\cal G}^{ab}_{k^+}(\underline{x};\underline{y})= \lim_{N\rightarrow +\infty} \theta(x^+\!-\!y^+)\, \int \left(\prod_{n=1}^{N-1} \textrm{d}^2 \z_{n}\right)\, \bigg(\frac{-i(k^+\!+\!i\epsilon)N}{2\pi (x^+\!-\!y^+)}\bigg)^N \nonumber\\
&&\hspace{3cm}\times \exp\bigg[\frac{i(k^+\!+\!i\epsilon)N}{2 (x^+\!-\!y^+)} \sum_{n=0}^{N-1} (\z_{n+1}\!-\!\z_{n})^2\bigg]\U^{ab}(x^+,y^+,\{\z_n\})
\, ,  \label{PathIntDef}
\end{eqnarray}
with $N$ being the number of discretization steps, $\z_{0}=\y$ and $\z_{N}=\x$. Here, $\U^{ab}(x^+,y^+,\{\z_n\})$ is the discretized Wilson line in the adjoint representation, defined as
\begin{eqnarray}
\U^{ab}(x^+,y^+,\{\z_n\})&=& {\cal P}_{+} \Bigg\{\prod_{n=0}^{N-1}  \exp \bigg[ig \frac{(x^+\!-\!y^+)}{N}   \Big({\cal A}^{-}(z^+_{n},\z_{n})\cdot T\Big)\bigg]\Bigg\}^{ab}
\, ,  \label{DiscrWilsonLine}
\end{eqnarray}
where
\begin{equation}
z^+_{n}=y^++\frac{n}{N}(x^+\!-\!y^+)\, .
\end{equation}
The background scalar propagator ${\cal G}^{ab}_{k^+}(\underline{x};\underline{y})$ defined in Eq.\eqref{PathIntDef} satisfies the following convolution relation:
\begin{equation}
\int \textrm{d}^2 \z\; {\cal G}^{ac}_{k^+}(\underline{x};\underline{z})\;  {\cal G}^{cb}_{k^+}(\underline{z};\underline{y})= {\cal G}^{ab}_{k^+}(\underline{x};\underline{y})\;\; \textrm{for}\;\; x^+>z^+>y^+\label{PathIntIteration}\, .
\end{equation}
In the absence of background field ${\cal A}^{-}$, the background scalar propagator reduces to
\begin{equation}
{\cal G}_{k^+}(\underline{x};\underline{y})_{ab}\equiv \delta_{ab}\;  {\cal G}_{0,k^+}(\underline{x};\underline{y})\, ,
\end{equation}
where the free scalar propagator ${\cal G}_{0,k^+}(\underline{x};\underline{y})$ can be written as
\begin{eqnarray}
{\cal G}_{0,k^+}(\underline{x};\underline{y})&=&  \int \frac{\textrm{d}^2 \k}{(2\pi)^2}\, e^{i \k\cdot (\x\!-\! \y)} \int \frac{\textrm{d} k^-}{2\pi}\,  e^{-i k^- (x^+\!-\! y^+)} \;\; \tilde{G}_{0,R}(k)
\label{FreeScalMom}\\
&=& \theta(x^+\!-\!y^+)\, \int \frac{\textrm{d}^2 \k}{(2\pi)^2}\, e^{i \k\cdot (\x\!-\! \y)}\; e^{-\frac{i(x^+\!-\!y^+)\k^2}{2(k^+\!+\!i\epsilon)}}\label{FreeScalMixMom}\\
&=& -\frac{i k^+}{2\pi}\; \frac{\theta(x^+\!-\!y^+)}{(x^+\!-\!y^+)}\; e^{\frac{i k^+(\x\!-\! \y)^2}{2 (x^+\!-\!y^+)}}
\,   \label{FreeScalMix}
\end{eqnarray}
from the usual expression of the retarded free scalar propagator in momentum space,
\begin{equation}
\tilde{G}_{0,R}(k)=\frac{i}{(k^2+2(k^+\!+\!k^-)i\epsilon)}=\frac{i}{2 (k^+\!+\!i\epsilon)(k^-- \frac{\k^2}{2(k^+\!+\!i\epsilon)} +i\epsilon)}\, .\label{ScalPropMom}
\end{equation}
\subsection{Calculation of the next-to-eikonal corrections to the background propagator\label{sec:next2eik_propag}}
Our aim is to calculate the expression of the background propagator to next-to-eikonal accuracy starting from the solution of the scalar Green's equation which is given in Eq.\eqref{PathIntDef}. One way of representing the eikonal limit corresponds to take the $k^+\to\infty$ limit, while keeping the background field as well as other kinematical variables fixed. In this limit, the kinetic term dominates over the potential term in the path integral in Eq.\eqref{PathIntDef}. Therefore, it is natural to consider a generic path as a perturbation around the classical free path, i.e.
\beq
\label{zexp}
\z_n =\z^{\textrm{cl}}_n +\u_n\,,
\eeq
where the transverse positions $\z^{\textrm{cl}}_n$ sit on a straight line between the initial and final points, such that
\beq
\label{zclass}
\z^{\textrm{cl}}_n = \y + \frac{n}{N}(\x-\y)\,,
\eeq
and the perturbation $\u_n$ satisfies the boundary conditions $\u_0=\u_N=0$. Rewriting Eq.\eqref{PathIntDef} as a path integral over the perturbation $\u_n$, one gets
\begin{eqnarray}
&&\hspace{-1.2cm}{\cal G}^{ab}_{k^+}(\underline{x};\underline{y})= \theta(x^+\!-\!y^+)\, \exp\Biggl[ \frac{i k^+\left(\x-\y\right)^2}{2(x^+-y^+)}\Biggr] \lim_{N\rightarrow +\infty} \int \left(\prod_{n=1}^{N-1} \textrm{d}^2 \u_n\right)\, \bigg[\frac{-i k^+ N}{2\pi (x^+\!-\!y^+)}\bigg]^N \nonumber\\
& &\times \exp\bigg[\frac{i k^+ N}{2 (x^+\!-\!y^+)} \sum_{n=0}^{N-1} (\u_{n+1}\!-\u_{n})^2\bigg]\U^{ab}\Big(x^+,y^+,\{\z^{\textrm{cl}}_n +\u_n\}\Big)
\, .  \label{PathIntDef-2}
\end{eqnarray}
%
The overall phase factor in Eq.\eqref{PathIntDef-2} is exactly the same phase factor that appears in the definition of the free propagator, Eq.\eqref{FreeScalMix}. Therefore, it is natural to factor out the free scalar propagator from the background propagator and rewrite Eq.\eqref{PathIntDef-2} as
\beq
\label{Gfact}
\G^{ab}_{k^+}(\underline{x};\underline{y})=\G_{0, k^+}(\underline{x};\underline{y}) \;\; {\cal R}^{ab}_{k^+}(\underline{x};\underline{y})\,,
\eeq
where $ {\cal R}_{k^+}(\underline{x};\underline{y})$ is given by
\begin{eqnarray}
&&\hspace{-2.6cm}{\cal R}^{ab}_{k^+}(\underline{x};\underline{y})= 2\pi i\frac{(x^+-y^+)}{k^+} \lim_{N\rightarrow +\infty} \int \left(\prod_{n=1}^{N-1} \textrm{d}^2 \u_n\right)\, \bigg[\frac{-i k^+ N}{2\pi (x^+\!-\!y^+)}\bigg]^N \nonumber\\
& &\times \exp\bigg[\frac{i k^+ N}{2 (x^+\!-\!y^+)} \sum_{n=0}^{N-1} (\u_{n+1}\!-\u_{n})^2\bigg]\U^{ab}\Big(x^+,y^+,\{\z^{\textrm{cl}}_n +\u_n\}\Big)
\, .  \label{R1}
\end{eqnarray}
Equivalently, one can rearrange Eq.\eqref{R1} into
\beq
\label{R2}
&&\hspace{-1cm}{\cal R}^{ab}_{k^+}(\underline{x};\underline{y})= 2\pi i\frac{(x^+-y^+)}{k^+} \lim_{N\rightarrow +\infty} \int \left(\prod_{n=1}^{N-1} \textrm{d}^2 \u_n\right)\P_+\prod_{n=0}^{N-1}\Bigg\{\G_{0, k^+}\Big(z^+_{n+1},\u_{n+1};z^+_{n},\u_{n} \Big)\nonumber\\
&&\hspace{4cm}\times\exp\Bigg[\frac{(x^+-y^+)}{N}\;igT\cdot {\cal A}^- \Big(z^+_n,\z^{\textrm{cl}}_n +\u_n\Big)\Bigg]\Bigg\} \ ,
\eeq
separating the pieces associated with each discretization step. The contributions at each step factorize into a free scalar propagator and a gauge link, corresponding respectively to the kinetic and potential contributions.

\subsubsection{Expansion around the free classical path}

As argued previously, one needs to make an expansion around the free classical path, and thus treat the transverse fluctuations $\u_n$ as small in each of the gauge links in Eq. \eqref{R2}.
Let us first consider such an expansion at the level of an individual gauge link. For this purpose, it is convenient to use the Lie product formula

\beq
\label{LieProduct}
e^{A+B}=\lim_{l\rightarrow+\infty}  \bigg( e^{\frac{A}{l}}\, \cdot\, e^{\frac{B}{l}} \bigg)^l\, ,
\eeq
valid for two generic square matrices $A$ and $B$.  For small $B$, $e^{\frac{B}{l}}=1+O(B)$, so that the Lie product formula gives
\beq
\label{LieProduct2}
&&\hspace{-1cm} e^{A+B}=\lim_{l\rightarrow+\infty}  \Bigg\{e^{A} + \sum_{j=1}^{l}    e^{\frac{j\, A}{l}}\, \bigg(e^{\frac{B}{l}}-1\bigg)\, e^{\frac{(l-j)\, A}{l}}\nonumber\\
&&\hspace{0.5cm}  +\sum_{j=1}^{l\!-\!1}\sum_{i=1}^{l\!-\!j}    e^{\frac{j\, A}{l}}\, \bigg(e^{\frac{B}{l}}-1\bigg)\,e^{\frac{i\, A}{l}}\, \bigg(e^{\frac{B}{l}}-1\bigg)\, e^{\frac{(l-j-i)\, A}{l}} +O\Big(B^3\Big)
\Bigg\}\, .
\eeq
Thanks to the large $l$ limit, each sum can be written as an integral, and one obtains
\beq
\label{LieProduct3}
&&\hspace{-1cm} e^{A+B}=  e^{A} + \int_{0}^{1} \hspace{-0.2cm} ds \;   e^{s\, A}\, B\, e^{(1-s)\, A}
+\int_{0}^{1}\hspace{-0.2cm} ds \;\int_{0}^{1\!-\!s}\hspace{-0.4cm} du\;  e^{s\, A}\, B\,e^{u\, A}\, B\,  e^{(1-s-u)\, A} +O\Big(B^3\Big)
 \, .
\eeq
Note that in each factor $[\exp(B/l)-1]$, only the leading term $B/l$ survives in the large $l$ limit.

Eq. \eqref{LieProduct3} can be used to perform the expansion of each gauge link in Eq. \eqref{R2} for small $\u_n$. Identifying $A\propto {\cal A}_n^-$ and $B\propto  \delta{\cal A}_n^-(\u_n)$, and using the notations
\beq
  {\cal A}_n^-&=&{\cal A}^-(z^+_n,\z^{cl}_n)\; ,         \nonumber\\
 \delta{\cal A}_n^-(\u_n)&=&{\cal A}^-(z^+_n,\z^{cl}_n+\u_n)-{\cal A}^-(z^+_n,\z^{cl}_n)\, ,
\eeq
one gets
\beq
\label{1linkExp}
&&\hspace{-0.5cm} \exp\Bigg[\frac{(x^+\!-\!y^+)}{N}\;igT\cdot {\cal A}^- \Big(z^+_n,\z^{\textrm{cl}}_n +\u_n\Big)\Bigg]
=\exp\Bigg[\frac{(x^+\!-\!y^+)}{N}\;igT\cdot   \bigg({\cal A}_n^- +\delta{\cal A}_n^-(\u_n)\bigg) \Bigg]\nonumber\\
&&\hspace{-0.5cm}
=  e^{\left[\frac{(x^+\!-\!y^+)}{N}\;igT\cdot {\cal A}_n^-\right]}\nonumber\\
&&\hspace{0cm}
+ \int_{0}^{1} \hspace{-0.2cm} ds \;   e^{\left[\frac{s(x^+\!-\!y^+)}{N}\;igT\cdot {\cal A}_n^-\right]}\: \bigg\{\frac{(x^+\!-\!y^+)}{N}\;igT\cdot   \delta{\cal A}_n^-(\u_n) \bigg\}  \: e^{\left[\frac{(1\!-\!s)(x^+\!-\!y^+)}{N}\;igT\cdot {\cal A}_n^-\right]}\nonumber\\
&&\hspace{0cm}
+ \int_{0}^{1} \hspace{-0.2cm} ds \;\int_{0}^{1\!-\!s}\hspace{-0.4cm} du\;   e^{\left[\frac{s(x^+\!-\!y^+)}{N}\;igT\cdot {\cal A}_n^-\right]}\: \bigg\{\frac{(x^+\!-\!y^+)}{N}\;igT\cdot   \delta{\cal A}_n^-(\u_n) \bigg\}  \:e^{\left[\frac{u(x^+\!-\!y^+)}{N}\;igT\cdot {\cal A}_n^-\right]}\nonumber\\
&&\hspace{0.5cm}\times \bigg\{\frac{(x^+\!-\!y^+)}{N}\;igT\cdot   \delta{\cal A}_n^-(\u_n)  \bigg\} \:  e^{\left[\frac{(1\!-\!s\!-\!u)(x^+\!-\!y^+)}{N}\;igT\cdot {\cal A}_n^-\right]}+O\Big(\delta{\cal A}_n^-(\u_n)^3\Big)
 \, .
\eeq

The expansion of ${\cal R}^{ab}_{k^+}(\underline{x};\underline{y})$ for small fluctuations around the free classical path, or equivalently for small $\delta{\cal A}_n^-(\u_n)$, is obtained by inserting the expansion given in  Eq.\eqref{1linkExp} into Eq.\eqref{R2}. At zeroth order in $\delta{\cal A}_n^-(\u_n)$, using Eq.\eqref{PathIntIteration} to combine together the scalar propagators, one gets
\beq
\label{Roth1}
{\cal R}^{ab}_{k^+}(\underline{x};\underline{y})\biggr.\biggl|_{0^{th}}&=&= 2\pi i\frac{(x^+-y^+)}{k^+} \lim_{N\rightarrow +\infty} \int \left(\prod_{n=1}^{N-1} \textrm{d}^2 \u_n\right)\P_+\prod_{n=0}^{N-1}\Bigg\{\G_{0, k^+}\Big(z^+_{n+1},\u_{n+1};z^+_{n},\u_{n} \Big)\nonumber\\
&&\hspace{4cm}\times\exp\Bigg[\frac{(x^+-y^+)}{N}\;igT\cdot {\cal A}_n^-\Bigg]\Bigg\} \ ,\nonumber\\
&=&2\pi i\,\frac{(x^+-y^+)}{k^+}\;\G_{0,k^+}(x^+,\mathbf{0};y^+,\mathbf{0})\;\lim_{N\to\infty}\U^{ab}(x^+,y^+,\{\z^{cl}_n\})\; .
\eeq
In the continuum limit, this contribution becomes
\beq
\label{Roth}
{\cal R}^{ab}_{k^+}(\underline{x};\underline{y})\biggr.\biggl|_{0^{th}}=\U^{ab} (\underline{x};\underline{y})
\eeq
where the Wilson line $\U^{ab} (\underline{x};\underline{y})\equiv \U^{ab}(x^+,y^+,[\z^{cl}(z^+)])$ follows the free classical trajectory $\z^{cl}(z^+)$ defined as
\beq
\z^{cl}(z^+)=\y+\frac{(z^+-y^+)}{(x^+-y^+)}\: (\x-\y)\; .
\eeq

The higher order contributions in $\delta{\cal A}_n^-(\u_n)$  to ${\cal R}^{ab}_{k^+}(\underline{x};\underline{y})$ are obtained by inserting subleading terms from Eq.\eqref{1linkExp} at one or more discretization steps into Eq.\eqref{R2}.

For instance, inserting subleading terms at a single step $l$, one obtains the local contribution
\beq
\label{Rloc1}
&& \hspace{-0.8cm}{\cal R}^{ab}_{k^+}(\underline{x};\underline{y})\biggr.\biggl|_{loc}=2\pi i\,\frac{(x^+-y^+)}{k^+}\Bigg\{\lim_{N\to\infty}\sum_{l=0}^{N-1}\int d^2\u_l\,\G_{0,k^+}(x^+,\mathbf{0};z^+_l,\u_l)\,\U(x^+,z^+_l,\{\z_n^{cl}\})\,\nonumber\\
&&
\hspace{0cm}\times \Bigg[
 \int_{0}^{1} \hspace{-0.2cm} ds \;   e^{\left[\frac{s(x^+\!-\!y^+)}{N}\;igT\cdot {\cal A}_l^-\right]}\: \bigg(\frac{(x^+\!-\!y^+)}{N}\;igT\cdot   \delta{\cal A}_l^-(\u_l) \bigg) \: e^{\left[\frac{(1\!-\!s)(x^+\!-\!y^+)}{N}\;igT\cdot {\cal A}_l^-\right]}\nonumber\\
&&\hspace{0cm}
+ \int_{0}^{1} \hspace{-0.2cm} ds \;\int_{0}^{1\!-\!s}\hspace{-0.4cm} du\;   e^{\left[\frac{s(x^+\!-\!y^+)}{N}\;igT\cdot {\cal A}_l^-\right]}\: \bigg(\frac{(x^+\!-\!y^+)}{N}\;igT\cdot   \delta{\cal A}_l^-(\u_l) \bigg) \:e^{\left[\frac{u(x^+\!-\!y^+)}{N}\;igT\cdot {\cal A}_l^-\right]}\nonumber\\
&&\hspace{0.5cm}\times \bigg(\frac{(x^+\!-\!y^+)}{N}\;igT\cdot   \delta{\cal A}_l^-(\u_l)  \bigg) \:  e^{\left[\frac{(1\!-\!s\!-\!u)(x^+\!-\!y^+)}{N}\;igT\cdot {\cal A}_l^-\right]}+O\Big(\delta{\cal A}_l^-(\u_l)^3\Big)
\Bigg]\nonumber\\
&&
\hspace{4cm}\times
\G_{0,k^+}(z^+_l,\u_l;y^+,\mathbf{0})\,\U(z^+_l,y^+,\{\z_n^{cl}\})\Bigg\}^{ab}\!\!\! .
\eeq
In the continuum limit we have
\beq
\frac{1}{N}\sum_{l=0}^{N-1}\to\int_{y^+}^{x^+}\frac{dz^+}{x^+-y^+} \; .
\eeq
Hence, only the term of order $\delta{\cal A}_l^-(\u_l)$ in ${\cal R}^{ab}_{k^+}(\underline{x};\underline{y})\biggr.\bigl|_{loc}$  survives in the continuum limit. Moreover, the integral over $s$ becomes trivial. Then, one gets
\beq
\label{Rloc2}
&& \hspace{-0.8cm}{\cal R}^{ab}_{k^+}(\underline{x};\underline{y})\biggr.\biggl|_{loc}=2\pi i\,\frac{(x^+\!-\!y^+)}{k^+}\Bigg\{\int_{y^+}^{x^+}\hspace{-0.3cm}dz^+\int d^2\u\,\G_{0,k^+}(x^+,\mathbf{0};z^+,\u)\,\U(\underline{x};z^+,\u)\,\nonumber\\
&&
\hspace{-1cm}\times \bigg[igT\cdot  \left( {\cal A}^-(z^+,\z^{cl}(z^+)+\u)-{\cal A}^-(z^+,\z^{cl}(z^+))\right) \bigg] \: 
\G_{0,k^+}(z^+,\u;y^+,\mathbf{0})\,\U(z^+,\u;\underline{y})\Bigg\}^{ab}\!\!\! .
\eeq
Finally, one has to Taylor-expand the expression inside the square bracket in Eq.\eqref{Rloc2}, and integrate over $\u$ order by order.
Thanks to the formulae
\beq
&&\hspace{-0.8cm}
\int d^2\u\, \G_{0,k^+}(x^+,\mathbf{0};z^+,\u)\,\u^i\, \G_{0,k^+}(z^+,\u;y^+,\mathbf{0})=0 \; ,
\nonumber\\
&&\hspace{-0.8cm}
\int d^2\u\, \G_{0,k^+}(x^+,\mathbf{0};z^+,\u)\, \u^i\, \u^j\, \G_{0,k^+}(z^+,\u;y^+,\mathbf{0})=\frac{\delta^{ij}}{2\pi}\frac{(x^+\!-\!z^+)(z^+\!-\!y^+)}{(x^+\!-\!y^+)^2}\; ,
\eeq
one finds
\beq
\label{Rloc}
&& \hspace{-0.8cm}{\cal R}^{ab}_{k^+}(\underline{x};\underline{y})\biggr.\biggl|_{loc}=i\frac{(x^+\!-\!y^+)}{2k^+}\int_{y^+}^{x^+}dz^+
\frac{(x^+\!-\!z^+)(z^+\!-\!y^+)}{(x^+\!-\!y^+)^2}
\Bigg\{\U\left(x^+,\x;z^+,\z^{cl}(z^+)\right)\nonumber\\
&&\hspace{-0.8cm}\times\biggl[igT\cdot \partial^2_{\z}{\cal A}^-\left(z^+,\z^{cl}(z^+)\right)
\biggr]\U\left(z^+,\z^{cl}(z^+);y^+,\y\right)\Bigg\}^{ab}
\hspace{-0.3cm}+ O\left(\left(\frac{(x^+\!-\!y^+)}{k^+}\, \partial^2_{\perp}\right)^2 \right).
\eeq
Higher order corrections, suppressed by additional powers of $(x^+\!-\!y^+)/k^+$, indeed come with additional transverse derivatives of the background field ${\cal A}^-$, generically denoted by $\partial_{\perp}$, defining a dimensionless expansion parameter.

${\cal R}^{ab}_{k^+}(\underline{x};\underline{y})$ also receives non-local contributions, constructed by inserting  subleading terms from Eq.\eqref{1linkExp} at more than one step into Eq.\eqref{R2}. Following the same reasoning as for the local contribution, one finds that corrections with more than one power of $\delta{\cal A}_l^-(\u_l)$ at any step $l$ vanish in the continuum limit. Then, non-local contributions to
${\cal R}^{ab}_{k^+}(\underline{x};\underline{y})$ can be written as straightforward generalizations to Eq.\eqref{Rloc2}. In particular, the bilocal contribution reads
\beq
\label{Rbiloc2}
&& \hspace{-0.4cm}{\cal R}^{ab}_{k^+}(\underline{x};\underline{y})\biggr.\biggl|_{biloc}=2\pi i\,\frac{(x^+\!-\!y^+)}{k^+}\Bigg\{\int_{y^+}^{x^+}\hspace{-0.3cm}dz^+
\int_{y^+}^{z^+}\hspace{-0.3cm}dw^+\int d^2\u\, \int d^2\v\,\nonumber\\
&&
\hspace{0cm}\times
\G_{0,k^+}(x^+,\mathbf{0};z^+,\u)\,\U(\underline{x};z^+,\u)\, \bigg[igT\!\cdot\!  \left( {\cal A}^-(z^+,\z^{cl}(z^+)+\u)\!-\!{\cal A}^-(z^+,\z^{cl}(z^+))\right) \bigg] \:\nonumber\\
&&
\hspace{-0.2cm}\times  \G_{0,k^+}(z^+,\u;w^+,\v)\,\U(z^+,\u;w^+,\v)\,\bigg[igT\!\cdot \! \left( {\cal A}^-(w^+,\z^{cl}(w^+)+\v)\!-\!{\cal A}^-(w^+,\z^{cl}(w^+))\right) \bigg]\nonumber\\
&&
\hspace{4cm}\times
\G_{0,k^+}(w^+,\v;y^+,\mathbf{0})\,\U(w^+,\v;\underline{y})\Bigg\}^{ab}\!\!\! .
\eeq
After Taylor-expanding at small $\u$ and $\v$, and performing transverse integrations using the formula
\beq
&&\hspace{-0.5cm}\int\, d^2\u\, d^2\v\,\G_{0,k^+}(x^+,\mathbf{0};z^+,\u)\, \u^i\, \G_{0,k^+}(z^+,\u;w^+,\v) \, \v^j\,\G_{0,k^+}(w^+,\v;y^+,\mathbf{0}) \nonumber\\
&& \hspace{2cm}=\frac{\delta^{ij}}{2\pi}\,\frac{(x^+-z^+)(w^+-y^+)}{(x^+-y^+)^2}\; ,
\eeq
one obtains
\beq
\label{Rbiloc}
&&\hspace{-0.4cm}{\cal R}^{ab}_{k^+}(\underline{x};\underline{y})\biggr.\biggl|_{biloc}
=i\frac{(x^+\!-\!y^+)}{k^+}\int_{y^+}^{x^+}\hspace{-0.2cm}dz^+
\int_{y^+}^{z^+}\hspace{-0.2cm}dw^+
\frac{(x^+\!-\!z^+)(w^+\!-\!y^+)}{(x^+\!-\!y^+)^2}
\Bigg\{\U\left(x^+,\x;z^+,\z^{cl}(z^+)\right)\nonumber\\
&&\hspace{-0.3cm}\times
\biggl[igT\cdot\partial_{\z^i}A^-\left(z^+,\z^{cl}(z^+)\right)\biggr]\,
\U\left(z^+,\z^{cl}(z^+);w^+,\z^{cl}(w^+)\right)
\biggl[igT\!\cdot\!\partial_{\z^i}A^-\left(w^+,\z^{cl}(w^+)\right)\biggr]\nonumber\\
&&\hspace{3cm}\times\,\U\left(w^+,\z^{cl}(w^+);y^+,\y\right)\Bigg\}^{ab} \hspace{-0.3cm}+ O\left(\left(\frac{(x^+\!-\!y^+)}{k^+}\, \partial^2_{\perp}\right)^2 \right) .
\eeq

Other non-local contributions to ${\cal R}^{ab}_{k^+}(\underline{x};\underline{y})$ appear only at higher orders in the large $k^+$ expansion. All in all, the ratio ${\cal R}^{ab}_{k^+}(\underline{x};\underline{y})$ of background and free propagators can be expanded at large $k^+$ as
\beq
\label{Rquex}
{\cal R}^{ab}_{k^+}(\underline{x};\underline{y})=\U^{ab} (\underline{x};\underline{y})+{\cal R}^{ab}_{k^+}(\underline{x};\underline{y})\biggr.\biggl|_{loc}+{\cal R}^{ab}_{k^+}(\underline{x};\underline{y})\biggr.\biggl|_{biloc}+ O\left(\left(\frac{(x^+\!-\!y^+)}{k^+}\, \partial^2_{\perp}\right)^2 \right)\, ,
\eeq
with the second and third terms on the right hand side given by Eqs.\eqref{Rloc} and \eqref{Rbiloc} respectively.

\subsubsection{Expansion around the initial transverse position}

The trajectory through the background field of a gluon with infinite $k^+$ is expected to follow a line of fixed transverse position, determined by the initial point. Indeed, the background propagator appears in practice in the calculation of the observables through its Fourier transform as
\beq
\label{FTofG}
&&\hspace{-1cm}\int d^2\x \; e^{-i\k\cdot\x}\;{\cal G}^{ab}_{k^+}(\underline{x};\underline{y})= -\frac{i k^+}{2\pi}\; \frac{\theta(x^+\!-\!y^+)}{(x^+\!-\!y^+)}\; \int d^2\x \; e^{-i\k\cdot\x}\;e^{\frac{i k^+(\x\!-\! \y)^2}{2 (x^+\!-\!y^+)}}{\cal R}^{ab}_{k^+}(\underline{x};\underline{y})\; .
\eeq
Thanks to the change of variable $\x\mapsto\h$, with 
\beq
\label{cov}
\x=\y+\frac{(x^+\!-\!y^+)}{k^+}\k+\sqrt{\frac{(x^+\!-\!y^+)}{k^+}}\,\h \; \; ,
\eeq
Eq.\eqref{FTofG} can be written as
\beq
\label{rescaledFTofG}
&&\hspace{-2cm}\int d^2\x \; e^{-i\k\cdot\x}\;{\cal G}^{ab}_{k^+}(\underline{x};\underline{y})= \theta(x^+\!-\!y^+)\;e^{-i\k\cdot\y}\;e^{-ik^-(x^+-y^+)}\nonumber\\
&&\times\int \frac{d^2\h}{2\pi i}\;e^{i\frac{\h^2}{2}}\;{\cal R}^{ab}_{k^+}\left(x^+,\y+\frac{(x^+\!-\!y^+)}{k^+}\k+\sqrt{\frac{(x^+\!-\!y^+)}{k^+}}\,\h;y^+,\y\right) \; .
\eeq
Hence, in addition to the expansion of ${\cal R}^{ab}_{k^+}(\underline{x};\underline{y})$ around the free classical trajectory at fixed end points performed in the previous subsection, one should also expand ${\cal R}^{ab}_{k^+}(\underline{x};\underline{y})$ for small $\x\!-\!\y$, since it is clear from Eqs.\eqref{cov} and \eqref{rescaledFTofG} that $\x\!-\!\y$ is parametrically small in the large $k^+$ limit.

There are three contributions at the order $1/k^+$ in Eq.\eqref{rescaledFTofG} at large $k^+$. The first one arises from the Taylor expansion of the term
$\U^{ab} (\underline{x};\underline{y})$ from Eq.\eqref{Rquex} to first order around $\x=\y$. In this contribution, the term proportional to $1/\sqrt{k^+}$ is linear in $\h$ and thus vanishes upon integration over $\h$, so that only the term proportional to $\k/k^+$ survives. The second contribution arises from the second order term in the Taylor expansion around $\x=\y$  of  $\, \U^{ab} (\underline{x};\underline{y})$ from Eq.\eqref{Rquex}. In that case, only the term quadratic in $\h$ is relevant at order $1/k^+$. Finally, the remaining terms from Eq.\eqref{Rquex} are already of order $1/k^+$. Hence, we should just make the substitution $\x\rightarrow\y$ in the expressions
of the local and bilocal contributions to ${\cal R}^{ab}_{k^+}(\underline{x};\underline{y})$, from the expansion around the free classical path,
which are given in Eqs.\eqref{Rloc} and \eqref{Rbiloc} respectively.

The last missing ingredient in order to write the $1/k^+$ contributions in Eq.\eqref{rescaledFTofG} is then the Taylor expansion of $\U^{ab} (\underline{x};\underline{y})$ to second order around $\x=\y$. This can be calculated in the discretized form of the Wilson line. The result reads
\beq
\label{TaylorU}
&&\hspace{-0.5cm} \U^{ab} (\underline{x};\underline{y}) =  \U^{ab} (x^+,y^+,\y)\nonumber\\
&&+ (\x^{i}\!-\!\y^{i}) \int_{y^+}^{x^+} \hspace{-0.3cm} dz^+\! \left(\frac{z^+\!-\!y^+}{x^+\!-\!y^+}\right)
\Big\{ \U (x^+,z^+,\y)  \left[igT\cdot\partial_{\y^i}A^-\left(z^+,\y\right)\right] \U (z^+,y^+,\y) \Big\}^{ab}\nonumber\\
&&+\frac{1}{2}(\x^{i}\!-\!\y^{i})(\x^{j}\!-\!\y^{j}) \int_{y^+}^{x^+} \hspace{-0.3cm} dz^+\! \left(\frac{z^+\!-\!y^+}{x^+\!-\!y^+}\right)^2\nonumber\\
&& \hspace{1cm} \times
\Big\{ \U (x^+,z^+,\y)  \left[igT\cdot\partial_{\y^i}\partial_{\y^j}A^-\left(z^+,\y\right)\right] \U (z^+,y^+,\y) \Big\}^{ab}\nonumber\\
&&+(\x^{i}\!-\!\y^{i})(\x^{j}\!-\!\y^{j}) \int_{y^+}^{x^+} \hspace{-0.3cm} dz^+\!\int_{y^+}^{z^+} \hspace{-0.3cm} dw^+\!  \frac{(z^+\!-\!y^+)(w^+\!-\!y^+)}{(x^+\!-\!y^+)^2}  \Big\{ \U (x^+,z^+,\y)\nonumber\\
&& \times    \left[igT\cdot\partial_{\y^i}A^-\left(z^+,\y\right)\right] \U (z^+,w^+,\y)  \left[igT\cdot\partial_{\y^j}A^-\left(w^+,\y\right)\right]\U (w^+,y^+,\y) \Big\}^{ab} ,
\eeq
using the notation $ \U^{ab} (x^+,y^+,\y)\equiv  \U^{ab} (x^+,\y;y^+,\y)$ for simplicity.
In order to write the final result one needs to integrate over $\h$. This can be performed thanks to the following formulae.
\beq
\label{hinteg}
\int \frac{d^2\h}{2\pi i}\;e^{i\frac{\h^2}{2}}=1\;,\hspace{0.7cm}
\int \frac{d^2\h}{2\pi i}\;e^{i\frac{\h^2}{2}}\h^i\;=\; 0 \hspace{0.7cm}\text{and}\hspace{0.4cm}
\int \frac{d^2\h}{2\pi i}\;e^{i\frac{\h^2}{2}}\h^i\h^j\;=\;i\;\delta^{ij}\; .
\eeq
Finally, collecting all the $1/k^+$ contributions one can rewrite Eq.\eqref{rescaledFTofG} as
\beq
\label{finalexpanded}
&&\hspace{-1cm}\int d^2\x \; e^{-i\k\cdot\x}\;{\cal G}^{ab}_{k^+}(\underline{x};\underline{y})= \theta(x^+\!-\!y^+)\;e^{-i\k\cdot\y}\;e^{-ik^-(x^+-y^+)}\Bigg\{ \U (x^+,y^+,\y)\;\nonumber\\
&&\hspace{0.5cm}+\;\frac{(x^+\!-\!y^+)}{k^+}\k^i\;\U ^i_{(1)}(x^+,y^+,\y)\,+\;i\frac{(x^+\!-\!y^+)}{2k^+}\;\U _{(2)}(x^+,y^+,\y)\nonumber\\
&&\hspace{5cm} \;+\; O\left(\left(\frac{(x^+\!-\!y^+)}{k^+}\, \partial^2_{\perp}\right)^2 \right)\Bigg\}^{ab} \; ,
\eeq
with the \emph{decorated Wilson lines} $\U^{i,ab}_{(1)}(x^+,y^+,\y)$ and $\U^{ab}_{(2)}(x^+,y^+,\y)$  defined as
\beq
\label{decoratedWilson1A}
&&\hspace{-1.5cm}\U^{i,ab}_{(1)}(x^+,y^+,\y)=\int_{y^+}^{x^+}\!\!\!\!  dz^+\!\left(\frac{z^+\!-\!y^+}{x^+\!-\!y^+}\right)\bigg\{\U(x^+,z^+,\y)\nonumber\\
&&\hspace{4cm}\times
\bigl[igT\cdot\partial_{\y^i}A^-(z^+,\y)\bigr]\U(z^+,y^+,\y)\bigg\}^{ab}\;
\eeq
and
\beq
\label{decoratedWilson2A}
&&\hspace{-0.3cm}\U^{ab}_{(2)}(x^+,y^+,\y)=\!\!
\int_{y^+}^{x^+}\!\!\!\!\!\!dz^+\left(\frac{z^+\!-\!y^+}{x^+\!-\!y^+}\right)\bigg\{\U(x^+,z^+,\y)\bigl[igT\cdot \partial^2_{\y}A^-(z^+,\y)
\bigr]\U(z^+,y^+,\y)\bigg\}^{ab}\nonumber\\
&&\hspace{1.2cm}+2\int_{y^+}^{x^+}\!\!\!\!\!\!dz^+\int_{y^+}^{z^+}\!\!\!\!\!\!dw^+
\left(\frac{w^+\!-\!y^+}{x^+\!-\!y^+}\right)\bigg\{\U(x^+,z^+,\y)\bigl[igT\cdot\partial_{\y^i}A^-(z^+,\y)\bigr]\nonumber\\
&&\hspace{4.2cm}\times\U(z^+,w^+,\y)\bigl[igT\cdot\partial_{\y^i}A^-(w^+,\y)\bigr]\,\U(w^+,y^+,\y)\bigg\}^{ab} \!\! .
\eeq
Note that  $\U^{ab}_{(2)}(x^+,y^+,\y)$ has a contribution that originates from the expansion around the free classical path and another contribution associated with the quadratic terms in $(\x-\y)$ from the Taylor expansion of the Wilson line given in Eq.\eqref{TaylorU}.

After some algebra, one can further simplify the expressions for $\, \U^{i,ab}_{(1)}(x^+,y^+,\y)$ and $\, \U^{ab}_{(2)}(x^+,y^+,\y)$ as
given in Eqs. \eqref{decoratedWilson1} and \eqref{decoratedWilson2}.

\section{Single inclusive gluon production in pA beyond the eikonal approximation}
\label{sec:kt_fact_beyond_eikonal}

As an example of application of the eikonal expansion performed in the previous section at the level of the gluon background propagator, let us consider the case of single inclusive gluon production in proton-nucleus collisions at high energy.

When the gluon is produced at mid-rapidity, this process is described within the CGC, assuming large rapidity separations between the produced gluon and both the proton and the nucleus\footnote{By contrast, when the rapidity separation between the produced parton and the projectile is small, the so-called hybrid factorization \cite{Dumitru:2005gt,Altinoluk:2011qy,Chirilli:2012jd,Kang:2014lha} is obtained.}, and a $k_\perp$-factorized formula is obtained \cite{Kovchegov:1998bi,Kovchegov:2001sc,Dumitru:2001ux,Kharzeev:2003wz,Blaizot:2004wu}. In these studies, the coupling of both the proton and the nucleus to the produced gluon are treated in the eikonal approximation. Our aim is to include some of the corrections beyond the eikonal approximation: the corrections associated with the finite longitudinal extent of the nucleus or equivalently with the time evolution of the partonic content of the proton while crossing the nucleus.  These should be the main next-to-eikonal corrections in the case of a large enough nucleus\footnote{General expressions for finite size targets were obtained in \cite{Wiedemann:2000za,Armesto:2013fca} but a systematic power expansion was never done.}.

\subsection{Semi-classical formalism\label{sec:semiclassical}}

In the CGC formalism, a highly boosted left-moving nucleus is usually described by a classical gluon shockwave ${\cal A}^{\mu}_a(x)=\delta^{\mu -}\: \delta(x^+)\: {\cal A}^{-}_a(\x)$ in the light-cone gauge $A^{+}_a=0$. That field has indeed a vanishing longitudinal width and no $x^-$ dependence in the limit of infinite boost.

Consider instead a background field
\beq
{\cal A}^{\mu}_a(x)=\delta^{\mu -}\: {\cal A}^{-}_a(x^+,\x)
\eeq
with a finite support  along the $x^+$ direction, from $x^+=0$ to $x^+=L^+$. In the case of a large nucleus, this should be the dominant finite-boost correction with respect to the usual gluon shockwave.

On the other hand, a highly boosted right-moving proton, considered as dilute, is described by a classical color current
\beq
j^{\mu}_a(x)=\delta^{\mu -}\: j^{+}_a(x)
\eeq
with zero width along $x^-$: $j^{+}_a(x)\propto\delta(x^-)$. This color current actually corresponds to the color distribution of the large-$x$ partons inside the proton, which appear as static color charges from the point of view of a low-$x$ gluon due to Lorentz dilation. Hence, in the CGC formalism, there is a particular $x=x_{\textrm{cut}}$ separating partons inside the proton into large-$x$ and low-$x$ ones.

Let us consider a proton-nucleus collision with a particular impact parameter $\B$, and choose the center of the nucleus as the reference point for the transverse plane, so that a generic point $\x$ in the transverse plane is at a distance $|\x\!-\!\B|$ from the center of the proton and at a distance $|\x|$ from the center of the nucleus.
Then, the color current $j^{+}_a(x)$ can be written as
\beq
j^{+}_a(x)=\delta(x^-)\; \U^{ab}(x^+,-\infty,\x)\;\;  \rho^{b}(\x\!-\!\B),
\eeq
where $\rho^{b}$ is the transverse density of color charges inside the proton before it reaches the nucleus, and $\U^{ab}(x^+,-\infty,\x)$ is the Wilson line implementing the color precession of these color charges in the background field ${\cal A}^{-}_a(x^+,\x)$ of the nucleus.

The single inclusive gluon spectrum for a pA collision with the impact parameter $\B$ reads
\beq
(2\pi)^3\, (2k^+)\, \frac{dN}{dk^+\, d^2\k} (\B)= \sum_{\lambda\, \textrm{phys.}}   \left\langle \left\langle  \left| {\cal M}^a_{\lambda}(\underline{k}, \B)\right|^2  \right\rangle_{p} \right\rangle_{A}\, ,
\label{Def_N}
\eeq
where $\lambda$,  $a$ and $\underline{k}=(k^+,\k)$ are the polarization, color and momentum of the produced real gluon,  and the ordering of the averages is irrelevant. Hereafter the summation over the color indices is kept implicit.
The cross-section for the single inclusive gluon production is then obtained by integrating over the impact parameter $\B$ as
\beq
\frac{d\sigma}{dk^+\, d^2\k}= \int d^2 \B \:\:  \frac{dN}{dk^+\, d^2\k} (\B)   \, .
\label{Def_cross-section}
\eeq
In the CGC effective theory, taking the expectation value of some operator in the nucleus state amounts to the classical statistical averaging $\langle\dots \rangle_{A}$ over the background field ${\cal A}^{-}_a$. The averaging $\langle\dots \rangle_{p}$ over the color charge density $\rho^a$ stands for taking the expectation value of some operator in the proton state.

At leading order in the coupling $g$, the amplitude for the production of a real gluon with momentum $\underline{k}=(k^+,\k)$, polarization $\lambda$ and color $a$ is given by the LSZ reduction formula
\beq
{\cal M}^a_{\lambda}(\underline{k}, \B)&=& \varepsilon_{\lambda}^{\mu *}  \int d^4x\; e^{i k\cdot x}\; \Box_{x} A_{\mu}^a(x)\ ,\label{LSZ_1}
\eeq
where $A_{\mu}^a(x)$ is the retarded\footnote{The demonstration given in Ref. \cite{Gelis:2006yv} that one can use retarded instead of Feynman propagators at tree level is valid for dense-dense collisions, in which the color charge densities of both colliding particles are in a classical representation \cite{Jeon:2004rk} and thus commute. In the case of dense-dilute collisions, one expect that the ordering of insertions of the color charge density $\rho^{a}$ of the projectile becomes relevant, potentially spoiling the demonstration from Ref. \cite{Gelis:2006yv}. However, for typical dense-dilute observables like the ones considered here, LO contributions come with exactly two insertions of $\rho^{a}$ (one in the amplitude and one in the complex conjugate amplitude), whose ordering do not matter due to the cyclicity of the trace corresponding to the projectile color average. Hence, one can still use retarded instead of Feynman propagators for typical dense-dilute observables at LO accuracy.} classical field \cite{Gelis:2006yv}. Since the nucleus is considered dense and the proton dilute, one has the power counting ${\cal A}^{-}_a(x^+,\x)={\cal O}(1/g)$ and $j^{+}_a(\underline{x})={\cal O}(g)$. Then, the classical field $A_{\mu}^a(x)$ appearing in the reduction formula, Eq.\eqref{LSZ_1}, has itself a perturbative expansion
\beq
A^{\mu}_a(x)={\cal A}^{\mu}_a(x)+ a^{\mu}_a(x) + {\cal O}(g^3)\, ,
\eeq
where
\beq
a^{\mu}_a(x)&=&-i \int d^4y\;  G^{\mu\nu}_{R}(x,y)_{ab}\; \: j^{b}_{\nu}(y)\nonumber\\
&=&-i \int d^4y\;  G^{\mu-}_{R}(x,y)_{ab}\;  \:  j^{+}_b(y)
\eeq
is the linearized field perturbation produced by the color current $j^{\mu}_a(x)$ on top of the background field ${\cal A}^{-}_a(x^+,\x)$.

In the light-cone gauge $A^+_a=0$, the polarization vectors satisfy $\varepsilon_{\lambda}^{+ *}=0$ for physical polarizations, so that $\varepsilon_{\lambda}^{\mu *}\, {\cal A}_{\mu}^a(x)=0$. Therefore, only the transverse components of the field disturbance $a^{\mu}_a(x)$ contribute to Eq.\eqref{LSZ_1} at leading order, which can be rewritten as
\beq
{\cal M}^a_{\lambda}(\underline{k}, \B)&=& - \varepsilon_{\lambda}^{i *}\:  (-2k^+ i) \lim_{x^+\rightarrow +\infty}  \int d^2\x  \int d x^-  \; e^{i k\cdot x}\; a^{i}_a(x)
\nonumber\\
&=&  \varepsilon_{\lambda}^{i *}\:  \lim_{x^+\rightarrow +\infty} e^{i k^- x^+} \int d^2\x\; e^{-i \k\cdot \x}  \int d^4 y\;
{\cal G}^{i-}_{k^+}(\underline{x};\underline{y})_{ab}\;\; j^{+}_b(y)\, ,
\label{M1}
\eeq
using the representation given in Eq.\eqref{FTPropMixed} for the gluon propagator.

The amplitude can be decomposed into three contributions as
\beq
{\cal M}^a_{\lambda}(\underline{k}, \B)&=& {\cal M}^a_{bef, \lambda}(\underline{k}, \B)+ {\cal M}^a_{in, \lambda}(\underline{k}, \B)+ {\cal M}^a_{aft, \lambda}(\underline{k}, \B)
\label{Mtot}
\eeq
in which the gluon is radiated by the current $j^{+}_a(x)$ respectively before, while, or after crossing the background field ${\cal A}^{-}_a(x^+,\x)$. Using the
expression for ${\cal G}^{i-}_{0,k^+}(\underline{x};\underline{y})$ given in Eq.\eqref{VectPerpMin2ScalMix} and the convolution property, Eq.\eqref{PathIntIteration}, of the background scalar propagator, one can separate the gluon propagator into parts in the vacuum and parts in the background field.
Then, plugging the expression for the free scalar propagator ${\cal G}_{0,k^+}$, Eq.\eqref{FreeScalMixMom}, it is straightforward to simplify the expressions for the three contributions ${\cal M}^a_{aft, \lambda}$, ${\cal M}^a_{in, \lambda}$ and  ${\cal M}^a_{bef, \lambda}$ as
\beq
{\cal M}^a_{aft,\lambda}(\underline{k}, \B)&=&  \varepsilon_{\lambda}^{i *}  \lim_{x^+\rightarrow +\infty} e^{i k^- x^+}\!\! \int d^2\x\; e^{-i \k\cdot \x} \!\int_{L^+}^{+\infty} d y^+ \int d^2\y\;\;
{\cal G}^{i-}_{0,k^+}(\underline{x};\underline{y})\nonumber\\
& &\qquad \quad \times \; \;   \U^{ac}(L^+,0,\y)     \;\; \rho^{c}(\y\!-\!\B)\nonumber\\
&=&  \varepsilon_{\lambda}^{i *}\: e^{i k^- L^+}  \int d^2\y\;\; e^{-i \k\cdot \y}\;\;\; 2i\, \frac{\k^i}{\k^2}
\;\;   \U^{ac}(L^+,0,\y)     \;\; \rho^{c}(\y\!-\!\B)\, ,
\label{outint}
\eeq
\beq
{\cal M}^a_{in,\lambda}(\underline{k}, \B)&=&  \varepsilon_{\lambda}^{i *}  \lim_{x^+\rightarrow +\infty} e^{i k^- x^+}\!\! \int d^2\x\; e^{-i \k\cdot \x} \!\int_{0}^{L^+} d y^+ \int d^2\y \int d^2\z'\;\;
{\cal G}_{0,k^+}(\underline{x},L^+,\z')\nonumber\\
& & \qquad \quad \times \;\; {\cal G}^{i-}_{k^+}(L^+,\z',\underline{y})_{ab}\;\;  \U^{bc}(y^+,0,\y)     \;\; \rho^{c}(\y\!-\!\B)\nonumber\\
&=&  \varepsilon_{\lambda}^{i *}\; e^{i k^- L^+} \int_{0}^{L^+} d y^+ \int d^2\z' \int d^2\y\;\; e^{-i \k\cdot \z'}\; \bigg[\frac{i}{k^+}\, \partial_{\y^i} {\cal G}^{ab}_{k^+}(L^+,\z',\underline{y})\bigg]
\nonumber\\
& & \qquad \quad \times \; \;  \U^{bc}(y^+,0,\y)     \;\; \rho^{c}(\y\!-\!\B)
\label{inint}
\eeq
and
\beq
{\cal M}^a_{bef,\lambda}(\underline{k}, \B)&=&  \varepsilon_{\lambda}^{i *}  \lim_{x^+\rightarrow +\infty} e^{i k^- x^+}\!\! \int d^2\x\; e^{-i \k\cdot \x} \!\int_{-\infty}^{0} d y^+ \int d^2\y \int d^2\z'\int d^2\z\;\;
{\cal G}_{0,k^+}(\underline{x},L^+,\z')\nonumber\\
& & \qquad \quad \times \;\; {\cal G}^{ac}_{k^+}(L^+,\z',0,\z)
\;\; {\cal G}^{i-}_{0,k^+}(0,\z,\underline{y})   \;\; \rho^{c}(\y\!-\!\B)\nonumber\\
&=&  \varepsilon_{\lambda}^{i *}\; e^{i k^- L^+}   \int d^2\z'  \int d^2\z\;\; e^{-i \k\cdot \z'}\; {\cal G}^{ac}_{k^+}(L^+,\z',0,\z)  \int d^2\y \;\; \rho^{c}(\y\!-\!\B)
\nonumber\\
& & \qquad \quad \times \; \;  \int \frac{d^2 \p}{(2\pi)^2} e^{i\p\cdot(\z\!-\!\y)}\;\;(-2i)\, \frac{\p^i}{\p^2}\, .
\label{befint}
\eeq
%
\subsection{Averaging over the projectile}

In the calculation of the single inclusive gluon cross section, Eq.\eqref{Def_cross-section}, one needs to evaluate the projectile average of color charge densities at equal momenta, $\left\langle \tilde{\rho}^{a}(\q)^{*}\: \tilde{\rho}^{b}(\q) \right\rangle_{p}$, where $\tilde{\rho}^{a}(\q)$ is defined by
\beq
\rho^{a}(\y-\B)= \int \frac{d^2 \q}{(2\pi)^2}\; e^{i \q\cdot(\y-\B)}\; \tilde{\rho}^{a}(\q)\, .
\label{Fourier_rho}
\eeq
This correlator $\left\langle \tilde{\rho}^{a}(\q)^{*}\: \tilde{\rho}^{b}(\q) \right\rangle_{p}$ is related to the unintegrated gluon distribution in the projectile with $\q$ being the transverse momentum of the gluon. For instance, in light-front quantization, one can formally decompose the physical one-proton state on a partonic Fock state basis, where each Fock state contains only partons with $x>x_{\textrm{cut}}$. Then, in each element of that basis, one can calculate both the correlator of the color charge density of the large-$x$ partons as well as the number density of the Weizs\"acker-Williams lower-$x$ gluons radiated by those large-$x$ partons. Comparing the expressions obtained for these two quantities, one finds the relation
\beq
\left\langle \tilde{\rho}^{a}(\q)^{*}\: \tilde{\rho}^{b}(\q) \right\rangle_{p}&=&
\frac{\delta^{ab}}{N_c^2\!-\!1}\;\;  \frac{(2\pi)^3}{2}\;\; \q^2\;\; \varphi_p(\q;x_{\textrm{cut}})
\label{rhorho_vs_ugd}\, ,
\eeq
where the normalization of the unintegrated gluon distribution $\varphi_p(\q;x_{\textrm{cut}})$ is such that the usual integrated gluon distribution at low-$x$ is obtained as
\beq
xG(x,\mu_F^2) &=& \int d^2\q\;\; \theta(\mu_F^2\!-\!\q^2)\;\; \varphi_p(\q;x_{\textrm{cut}})\qquad   \textrm{for}\quad x<x_{\textrm{cut}}\, ,
\label{pdf_vs_ugd}
\eeq
with $\mu_F$ being the factorization scale for the collinear factorization.
The boost invariance of the classical Weizs\"acker-Williams gluon fields implies that these gluon distributions have no dependence on $x$  below $x_{\textrm{cut}}$, in the present setup.

However, it is natural to choose $x_{\textrm{cut}}=k^+/P^+$, where $P^+$ is the momentum of the proton, in order to resum BFKL leading logs into $\varphi_p$ in the regime $k^+\ll P^+$.

On the other hand, when calculating the single inclusive gluon spectrum, Eq.\eqref{Def_N}, one needs the correlator $\left\langle\rho^a(\yb-\B)\rho^b(\y-\B)\right\rangle_{p}$, which is related with the Wigner distribution,  ${\cal W}(\q,\b_p;x_{\textrm{cut}})$\footnote{Wigner functions are related both to the generalised parton distributions and to the unintegrated (or TMD) parton distributions \cite{Ji:2003ak,Belitsky:2003nz}.}, of Weizs\"acker-Williams gluons from the projectile via
\beq
{\cal W}(\q,\b_p;x_{\textrm{cut}})&=&
  \frac{2}{(2\pi)^3\; \q^2}\; \int d^2\r_p\;  e^{-i \q\cdot \r_p}\; \left\langle\rho^a\left(\b_p\!-\!\frac{\r_p}{2}\right)\; \rho^a \left(\b_p\!+\!\frac{\r_p}{2}\right)\right\rangle_{p}
\label{rhorho_vs_Wigner}\, ,
\eeq
where $\b_p$ is impact parameter of the gluon with respect to the projectile. The Wigner distribution is related to the aforementioned unintegrated gluon distribution as
\beq
\varphi_p(\q;x_{\textrm{cut}})&=&  \int d^2\b_p\; {\cal W}(\q,\b_p;x_{\textrm{cut}})
\label{ugd_vs_Wigner}\, .
\eeq
Hence, the required correlator reads
\beq
\hspace{-1cm}\left\langle\rho^a(\yb\!-\!\B)\rho^b(\y\!-\!\B)\right\rangle_{p}&=& \frac{\delta^{ab}}{N_c^2\!-\!1}\!\!  \int \frac{d^2\q}{(2\pi)^2}\; e^{-i \q\cdot (\y-\yb)}\;
  \frac{(2\pi)^3\q^2}{2}\;  {\cal W}\!\!\left(\q, \frac{\y\!+\!\yb}{2} \!-\!\B;x_{\textrm{cut}}\right) .       \label{rhorho_vs_Wigner_2}
\eeq
%

\subsection{Factorization in position space}

Now we can proceed in two different ways. Starting from the expressions for ${\cal M}^a_{bef,\lambda}(\underline{k}, \B)$, ${\cal M}^a_{aft,\lambda}(\underline{k}, \B)$ and ${\cal M}^a_{in,\lambda}(\underline{k}, \B)$ derived in section \ref{sec:semiclassical}, one can write the total scattering amplitude either in coordinate space or in  momentum space. In this section, we work in coordinate space and write down the factorized form of the square of the total scattering amplitude in coordinate space which can be then either plugged into the expression of single inclusive gluon spectrum, Eq.\eqref{Def_N}, or into the expression for the cross-section, Eq.\eqref{Def_cross-section}.

Let us start with ${\cal M}^a_{bef,\lambda}(\underline{k}, \B)$. After integrating over $\p$, the ``$bef$" contribution given in Eq.\eqref{befint} can be written as

\beq
\hspace{-1cm}{\cal M}^{a}_{bef,\lambda}(\underline{k},\B)&=&\varepsilon^{*i}_\lambda\,e^{i k^-L^+}\frac{1}{\pi}\int_{\y,\z}\hspace{-0.4cm}e^{-i\k\cdot\z}
\frac{(\z-\y)^i}{(\z-\y)^2}\int_{\z'}\hspace{-0.2cm}e^{i\k\cdot(\z-\z')}
\G^{ab}_{k^+}(L^+,\z';0,\z)\rho^b(\y-\B)\, .\label{bef-pos}
\eeq
Thanks to the identity
\beq
\frac{\k^i}{\k^2}=\frac{1}{2\pi i}\int_\z e^{i\k\cdot\z}\frac{\z^i}{\z^2}\, ,
\eeq
the ``$aft$" contribution, given in Eq.\eqref{outint}, reads
\beq
{\cal M}^{a}_{aft,\lambda}(\underline{k},\B)&=&-\varepsilon^{*i}_\lambda\,e^{i k^-L^+}\frac{1}{\pi}\int_{\y,\z}e^{-i\k\cdot\z}
\frac{(\z-\y)^i}{(\z-\y)^2}\,\U^{ab}(L^+,0,\y)\rho^b(\y-\B).\label{aft-pos}
\eeq
Both ``$bef$" and ``$aft$" contributions to the amplitude are now written in a factorized form. In order to have a similar factorization for the ``$in$" contribution to the amplitude, we first separate the $\y$ dependence from the background propagator with a $\delta$-function:
\beq
\G^{ab}_{k^+}(L^+,\z';y^+,\y)=\int_{\z}\delta^2(\z-\y)\,\G^{ab}(L^+,\z';y^+,\z)
\eeq
Using the fact that in two dimensions the Dirac delta function can be written as
\beq
\delta^2(\z-\y)=\frac{1}{2\pi}\partial^2_\z \ln (|\z-\y|)
\eeq
and performing integration by parts once to act with $\partial_{\y"}$ on the logarithm, the ``$in$'' contribution to the amplitude can be written as
\beq
\hspace{-0.5cm}{\cal M}^{a}_{in,\lambda}(\underline{k},\B)&=&\varepsilon^{*i}_\lambda e^{i k^-L^+}\frac{1}{2\pi}\frac{1}{ik^+}\int_{\y,\z,\z',y^+}\hspace{-1cm}e^{-i\k\cdot\z'}\partial^2_\z
\biggl[\frac{(\z-\y)^i}{(\z-\y)^2}\biggr]\G^{ab}_{k^+}(L^+,\z';y^+,\z)\U^{bc}(y^+,0,\y)\,\nonumber\\
&&\hspace{8cm}\times\,\rho^c(\y-\B).
\eeq
We can perform integration by parts twice to act with $\partial^2_\z$ on the background propagator and rewrite the final factorized form of the``$in$'' contribution as
\beq
\hspace{-0.5cm}{\cal M}^{a}_{in,\lambda}(\underline{k},\B)&=&\varepsilon^{*i}_\lambda e^{i k^-L^+}\frac{1}{2\pi}\frac{1}{i k^+}\int_{\y,\z}\hspace{-0.4cm}e^{-i\k\cdot\z}\frac{(\z-\y)^i}{(\z-\y)^2}\,\int_{\z',y^+}\hspace{-0.6cm}
e^{i\k\cdot(\z-\z')}\partial^2_\z\left[\G^{ab}_{k^+}(L^+,\z';y^+,\z)\right]\nonumber\\
&&\hspace{5cm}\times\;\U^{bc}(y^+,0,\y)\rho^c(\y-\B).\label{in-pos}
\eeq
Since we have all the contributions in a factorized form, the total amplitude now can be written as
\beq
\label{totM}
{\cal M}^{a}_{\lambda}(\underline{k},\B)&=&\varepsilon^{*i}_\lambda e^{i k^-L^+}\frac{1}{\pi}\int_{\y,\z}\hspace{-0.4cm}e^{-i\k\cdot\z}\frac{(\z-\y)^i}{(\z-\y)^2}\,\Biggl\{
\int_{\z'}e^{i\k\cdot(\z-\z')}\G_{k^+}(L^+,\z';0,\z)-\U(L^+,0,\y)\nonumber\\
&&\hspace{0.1cm}+\int_{\z',y^+}\hspace{-0.6cm}e^{i\k\cdot(\z-\z')}\frac{1}{2ik^+}\partial^2_\z
\left[\G_{k^+}(L^+,\z';y^+,\z)\right]\,\U(y^+,0,\y)
\Biggr\}^{ab}
\rho^b(\y-\B).
\eeq
Substituting the final expression for the total amplitude, Eq.\eqref{totM}, into the single inclusive gluon spectrum, Eq.\eqref{Def_N}, one gets

%
\beq
\hspace{0cm}&&(2\pi)^3\, (2k^+)\, \frac{dN}{dk^+\, d^2\k} (\B)\!\!= \!\frac{1}{\pi^2}\int_{\y,\z,\yb,\zb}\hspace{-0.6cm}\kappa (\yb,\zb,\y,\z)\Bigg\langle\Biggl\{
\int_{\zb'}\hspace{-0.2cm}e^{-i\k\cdot(\zb-\zb')} \G^{\dagger}_{k^+}(L^+,\zb';0,\zb)\nonumber\\
&&\hspace{1cm}-\U^{\dagger}(L^+,0,\yb)-\frac{1}{2ik^+}\int_{\zb',\bar{y}^+}\hspace{-0.6cm}e^{-i\k\cdot(\zb-\zb')} \U^{\dagger}(\bar{y}^+,0,\yb)\partial^2_{\zb}\left[\G_{k^+}^{\dagger}(L^+,\zb';\bar{y}^+,\zb)\right]
\Biggr\}^{ab}\nonumber\\
&&\hspace{1cm}\times\Biggl\{
\int_{\z'}\hspace{-0.2cm}e^{i\k\cdot(\z-\z')} \G_{k^+}(L^+,\z';0,\z)-\U(L^+,0,\y)\nonumber\\
&&\hspace{1.2cm}+\frac{1}{2ik^+}\int_{\z',y^+}e^{i\k\cdot(\z-\z')} \partial^2_{\z}\left[\G_{k^+}(L^+,\z';y^+,\z)\right]
\U(y^+,0,\y)
\Biggr\}^{bc}\Bigg\rangle_A\nonumber\\
&&\hspace{5.5cm}\times \left\langle\rho^a(\yb\!-\!\B)\rho^c(\y\!-\!\B)\right\rangle_{p}\ ,
\eeq
where $\kappa (\yb,\zb,\y,\z)$ is defined as
\beq
\kappa (\yb,\zb,\y,\z)=e^{i\k\cdot(\zb-\z)}\frac{(\zb-\yb)}{(\zb-\yb)^2}\cdot\frac{(\z-\y)}{(\z-\y)^2}\, .
\eeq
In the above expression one can express the color correlator of the projectile charge density in terms of the Wigner distribution via Eq.\eqref{rhorho_vs_Wigner_2}. Moreover, the single inclusive gluon cross section is straightforward to obtain as a function of the unintegrated gluon distribution by using Eq.\eqref{ugd_vs_Wigner}.

\subsection{$k_\perp$-factorization}

Alternatively, one can write the total amplitude, ${\cal M}^a_{\lambda}$, in a factorized form in momentum space. Using the Fourier transform
 of the color charge density, Eq.\eqref{Fourier_rho}, one writes
\beq
{\cal M}^a_{\lambda}(\underline{k}, \B)&=& \int \frac{d^2 \q}{(2\pi)^2}\; e^{-i \q\cdot\B}\;\;    \overline{{\cal M}}^{a b}_{\lambda}(\underline{k}, \q)\;\; \tilde{\rho}^{b}(\q)\, ,
\label{def_reduced_Ampl}
\eeq
defining a gluon-nucleus reduced amplitude $\overline{{\cal M}}^{a b}_{\lambda}(\underline{k}, \q)$. The three contributions ``$aft$",  ``$bef$" and ``$in$"  give at this level
\beq
\overline{{\cal M}}^{a b}_{aft,\lambda}(\underline{k}, \q)&=&  \varepsilon_{\lambda}^{i *}\: e^{i k^- L^+}\; i  \int d^2\y\;\; e^{i (\q-\k)\cdot \y}\;\;\; 2\, \frac{\k^i}{\k^2}
\;\;   \U^{ab}(L^+,0,\y)\, ,
\label{outred}
\eeq
\beq
\overline{{\cal M}}^{a b}_{in,\lambda}(\underline{k}, \q)&=&   \varepsilon_{\lambda}^{i *}\; e^{i k^- L^+}\; i
\int d^2\y\;\; e^{i \q\cdot \y}\; \frac{1}{k^+}\int_{0}^{L^+} \!\!\!\! d y^+ \int d^2\z'\;\; e^{-i \k\cdot \z'}\;
\nonumber\\
& & \qquad \quad \times \; \; \bigg[ \partial_{\y^i} {\cal G}^{ac}_{k^+}(L^+,\z',\underline{y})\bigg]\;\; \U^{cb}(y^+,0,\y)
\label{inred}
\eeq
and
\beq
\hspace{-0.8cm}
\overline{{\cal M}}^{a b}_{bef,\lambda}(\underline{k}, \q)&=&  \varepsilon_{\lambda}^{i *}\; e^{i k^- L^+}\; i
\int d^2\z\; e^{i \q\cdot \z}\;(-2)\, \frac{\q^i}{\q^2}
\int d^2\z'\; e^{-i \k\cdot \z'}\; {\cal G}^{ab} _{k^+}(L^+,\z',0,\z) \, .
\label{befred}
\eeq

Using the relation between the amplitude and the reduced amplitude, Eq.\eqref{def_reduced_Ampl}, the single inclusive gluon spectrum can be written as
\beq
&& (2\pi)^3\, (2k^+)\, \frac{dN}{dk^+\, d^2\k} (\B)=\int \frac{d^2 \q_1}{(2\pi)^2}\; \int \frac{d^2 \q_2}{(2\pi)^2}\;  \; e^{-i (\q_1-\q_2) \cdot\B}
\left\langle \tilde{\rho}^{c}(\q_2)^{*}\: \tilde{\rho}^{b}(\q_1) \right\rangle_{p}\nonumber\\
&& \hspace{6cm} \times \sum_{\lambda\, \textrm{phys.}} \left\langle  \overline{{\cal M}}^{a c}_{\lambda}(\underline{k}, \q_2)^{\dag}\;   \overline{{\cal M}}^{a b}_{\lambda}(\underline{k}, \q_1)  \right\rangle_{A}  ,
\label{dN_mom}
\eeq
where the notations $\q_1$ and  $\q_2$ are introduced to distinguish the momenta of the gluon in the amplitude and in the complex conjugate amplitude respectively.
Changing variables from $\q_1$ and  $\q_2$ to $\q=(\q_1+\q_2)/2$ and ${\bf \Delta}=(\q_1\!-\!\q_2)$ and using the definition of the Wigner distribution, Eq.\eqref{rhorho_vs_Wigner}, one obtains the expression
\beq
&& k^+\, \frac{dN}{dk^+\, d^2\k} (\B)=\int \frac{d^2 \q}{(2\pi)^2}\; \int \frac{d^2 \bf{\Delta}}{(2\pi)^2}\;  \int d^2 \b_p  \; e^{-i   \bf{\Delta}\cdot(\B+\b_p)}\;
{\cal W}(\q,\b_p;x_{\textrm{cut}})\;
\nonumber\\
&& \hspace{2cm} \times  \frac{\q^2}{4}\, \frac{1}{N_c^2\!-\!1}\, \sum_{\lambda\, \textrm{phys.}} \left\langle  \overline{{\cal M}}^{a b}_{\lambda}\left(\underline{k}, \q\!-\!\frac{\bf{\Delta}}{2} \right)^{\dag}\;   \overline{{\cal M}}^{a b}_{\lambda}\left(\underline{k}, \q\!+\!\frac{\bf{\Delta}}{2}\right)  \right\rangle_{A}  .
\label{dN_mom_2}
\eeq
As a remark, note that the combination $\B+\b_p$ appearing in Eq.\eqref{dN_mom_2} is the impact parameter of the gluon with respect to the center of the target.

In the literature, the single inclusive gluon spectrum is usually considered only in a large target approximation in order to simplify Eq.\eqref{dN_mom_2}. This approximation may be justified as follows. The impact parameter $\b_p$ of the gluon with respect to the proton projectile should be bounded by the radius of the proton. On the other hand, in the case of a large nuclear target, typical values of the impact parameter $\B$ between the projectile and the target are larger than the proton radius. This suggests to approximate the impact parameter $\B+\b_p$ of the gluon with respect to the target by  $\B$ in the phase in Eq.\eqref{dN_mom_2}. Then, the integration over $\b_p$ can be performed thanks to the relation between the unintegrated gluon distribution and the Wigner distribution given in Eq.\eqref{ugd_vs_Wigner}, and one obtains
\beq
&& k^+\, \frac{dN}{dk^+\, d^2\k} (\B) \simeq \int \frac{d^2 \q}{(2\pi)^2}\;  \varphi_p(\q;x_{\textrm{cut}})\; \frac{\q^2}{4}\, \frac{1}{N_c^2\!-\!1}\, \int \frac{d^2 \bf{\Delta}}{(2\pi)^2}\; e^{-i   \bf{\Delta}\cdot\B}\;
\nonumber\\
&& \hspace{3cm} \times   \sum_{\lambda\, \textrm{phys.}} \left\langle  \overline{{\cal M}}^{a b}_{\lambda}\left(\underline{k}, \q\!-\!\frac{\bf{\Delta}}{2} \right)^{\dag}\;   \overline{{\cal M}}^{a b}_{\lambda}\left(\underline{k}, \q\!+\!\frac{\bf{\Delta}}{2}\right)  \right\rangle_{A}  .
\label{dN_mom_large_A}
\eeq

In order to calculate the single inclusive gluon cross section, one should integrate the spectrum, Eq.\eqref{dN_mom_2}, over $\B$. The  $\B$ integration  forces the momentum transfer ${\bf \Delta}$ to be zero and finally the single inclusive gluon cross section reads
%
\beq
k^+\, \frac{d\sigma}{dk^+\, d^2\k}=\int \frac{d^2 \q}{(2\pi)^2}\;
 \varphi_p(\q;x_{\textrm{cut}})\;  \frac{ \q^2}{4}\;
\frac{1}{N_c^2\!-\!1}
\sum_{\lambda\, \textrm{phys.}} \left\langle  \overline{{\cal M}}^{a b}_{\lambda}(\underline{k}, \q)^{\dag}\;   \overline{{\cal M}}^{a b}_{\lambda}(\underline{k}, \q)  \right\rangle_{A} .
\label{cross-section_3}
\eeq
%

\section{Single inclusive gluon production in pA at next-to-eikonal accuracy}
\label{sec:kt_fact_next2eikonal}
The expressions for both the single inclusive gluon spectrum and the cross section, derived in section \ref{sec:kt_fact_beyond_eikonal}, involve the background scalar propagator $\G^{ab}_{k^+}(\underline{x};\underline{y})$, which has been studied in section \ref{sec:G_expansion}. The aim in this section is first to apply its eikonal expansion, Eq.\eqref{finalexpanded}, at the level of the amplitude and then  to obtain and discuss the next-to-eikonal corrections 
for both of the observables, as well as for related spin asymmetries.

\subsection{Next-to-eikonal corrections at the amplitude level}
\label{sec:next2eikonal_ampl}

We first consider the expansion of the amplitude in position space. The ``$aft$" contribution does not include the background scalar propagator, $\G^{ab}_{k^+}(\underline{x};\underline{y})$, and thus can be kept as written in Eq.\eqref{aft-pos}.
Using the eikonal  expansion of  $\G^{ab}_{k^+}(\underline{x};\underline{y})$ given in Eq.\eqref{finalexpanded}  and integrating over $z'$, the ``$bef$" contribution becomes
\beq
{\cal M}^{a}_{bef,\lambda}(\underline{k},\B)&=&\varepsilon^{*i}_\lambda\frac{1}{\pi}\int_{\y,\z}e^{-i\k\cdot\z}
\frac{(\z-\y)^i}{(\z-\y)^2}\bigg\{ \U(L^+,0;\z) +\frac{L^+}{k^+}\k^j\U^j_{(1)}(L^+,0;\z)\nonumber\\
&&\hspace{4.5cm}+\frac{i}{2}\frac{L^+}{k^+}\U_{(2)}(L^+,0;\z)\bigg\}^{ab}\rho^b(\y-\B)
\eeq
and the "$in$" contribution
\beq
\hspace{0cm}{\cal M}^{a}_{in,\lambda}(\underline{k},\B)&=&-\varepsilon^{*i}_\lambda\frac{1}{\pi}\frac{i}{2k^+}\int_{\y,\z,y^+}
\hspace{-0.5cm}e^{-i\k\cdot\z+ik^-y^+}
\frac{(\z-\y)^i}{(\z-\y)^2}\big[\partial_{\z^j}-i\k^j\big]^2\nonumber\\
&&\hspace{3.5cm}\times\{\U(L^+,y^+;\z)\U(y^+,0;\y)\}^{ab}\rho^b(\y-\B)\, .
\eeq
Adding all the contributions, we get the total amplitude as
\beq
\label{totMcoor}
&&\hspace{-0.5cm}{\cal M}^{a}_{\lambda}(\underline{k},\B)=\varepsilon^{*i}_\lambda\frac{1}{\pi}\int_{\y,\z}\hspace{-0.4cm}e^{-i\k\cdot\z}
\frac{(\z-\y)^i}{(\z-\y)^2}\bigg\{\U(L^+,0;\z)-\U(L^+,0;\y)+\frac{L^+}{k^+}\bigg[\k^j\U^j_{(1)}(L^+,0;\z)\nonumber\\
&&\hspace{-0.5cm}+\frac{i}{2}\U_{(2)}(L^+,0;\z)-i\frac{\k^2}{2}\U(L^+,0;\y)-\frac{i}{2}\int_{0}^{L^+}\frac{dy^+}{L^+}
\big[\partial_{\z^j}-i\k^j\big]^2\U(L^+,y^+;\z)\U(y^+,0;\y)\bigg]\bigg\}^{ab}\nonumber\\
&&\hspace{6cm}\times\rho^b(\y-\B)\hspace{0.4cm} + \hspace{0.3cm} O\left(\left(\frac{L^+}{k^+}\, \partial^2_{\perp}\right)^2 \right).
\eeq

By comparison, the same expansion can be also applied at the level of the gluon-nucleus reduced amplitude $\overline{{\cal M}}^{a b}_{\lambda}(\underline{k}, \q)$ in momentum space.  In this case the ``$bef$" contribution to $\overline{{\cal M}}^{a b}_{\lambda}(\underline{k}, \q)$ becomes
\beq
\hspace{-0.8cm}
\overline{{\cal M}}^{a b}_{bef,\lambda}(\underline{k}, \q)&=& i\; \varepsilon_{\lambda}^{i *}\;
\int d^2\z\; e^{i (\q-\k)\cdot \z}\;(-2)\, \frac{\q^i}{\q^2}\nonumber\\
&&\bigg\{\U(L^+,0;\z)+\frac{L^+}{k^+}\k^j\U_{(1)}^{j}(L^+,0;\z)+\frac{i}{2}\frac{L^+}{k^+}\U_{(2)}(L^+,0;\z)\bigg\}^{ab}\; .
\label{befredexp}
\eeq
Similarly, the ``$in$" contribution reads
\beq
\hspace{-0.8cm}
\overline{{\cal M}}^{a b}_{in,\lambda}(\underline{k}, \q)&=& i\; \varepsilon_{\lambda}^{i *}\;
\int d^2\y\; e^{i (\q-\k)\cdot \y}\;\bigg\{-2\, \frac{\k^i}{\k^2}\big(e^{ik^-L^+}-1\big)
\U(L^+,0;\y)\nonumber\\
&&+\frac{L^+}{k^+}\int_{0}^{L^+}\frac{dy^+}{L^+}e^{ik^-y^+}[\partial_{\y^i}\U(L^+,y^+;\y)]\U(y^+,0;\y)\bigg\}^{ab}\; .
\label{inredexp}
\eeq
The phase $e^{ik^-y^+}$ only affects the higher order terms in inverse powers of $k^+$, thus can be dropped. Then, one recognizes the decorated Wilson line $\U_{(1)}^{i}(L^+,0;\y)$ in the second line of Eq.\eqref{inredexp}. On top of this, the term in the first line combines with the ``$aft$" contribution, removing its phase $e^{ik^-L^+}$. Finally, the total gluon-nucleus reduced amplitude $\overline{{\cal M}}^{a b}_{\lambda}(\underline{k}, \q)$ reads
\beq
\hspace{0cm}
&&\overline{{\cal M}}^{a b}_{\lambda}(\underline{k}, \q)= i\; \varepsilon_{\lambda}^{i *}\;
\int d^2\y\; e^{-i (\k-\q)\cdot \y}\;\Bigg\{2\,\bigg[  \frac{\k^i}{\k^2} -\frac{\q^i}{\q^2}\bigg]
\U(L^+,0;\y)\nonumber\\
&&+\frac{L^+}{k^+}\bigg[\delta^{ij}-2\frac{\q^i}{\q^2}\k^j\bigg]\U_{(1)}^{j}(L^+,0;\y)
-i\frac{L^+}{k^+}\frac{\q^i}{\q^2}\U_{(2)}(L^+,0;\y)
+ O\left(\left(\frac{L^+}{k^+}\, \partial^2_{\perp}\right)^2 \right)
\Bigg\}^{ab}\; .
\label{totredexp}
\eeq
It is straightforward to check that Eq.\eqref{totredexp} and Eq.\eqref{totMcoor} are indeed equivalent, by using the relation between the amplitude and the reduced amplitude given in Eq.\eqref{def_reduced_Ampl}.
The total amplitude in momentum space, Eq.\eqref{totredexp}, has a more compact form than the one in position space, Eq.\eqref{totMcoor}. The momentum space expression, Eq.\eqref{totredexp}, will thus be used to study next-to-eikonal corrections to the observables.

\subsection{Eikonal expansion of the single inclusive gluon cross section}
\label{sec: next2eik_sigma}

The aim of this section is to investigate the influence of the next-to-eikonal corrections to the amplitude given in Eq.\eqref{totredexp} on the single inclusive gluon cross section. It is convenient to introduce the following notations\footnote{The quantities $S_A(\r,\b)$, ${\cal O}^j_{(1)}(\r,\b)$ and ${\cal O}_{(2)}(\r,\b)$ have rapidity divergences, and should thus depend on an extra parameter, depending on the regularization scheme. We keep this dependence implicit, in our calculation at strict LO accuracy. The dependence of $S_A(\r,\b)$ on this regulator is obtained at LL accuracy from the JIMWLK equation. However, the dependence of ${\cal O}^j_{(1)}(\r,\b)$ and ${\cal O}_{(2)}(\r,\b)$ on this regulator would be given by a more general evolution equation, see studies along this direction in Refs. \cite{Blaizot:2014bha,Iancu:2014kga}.}:
c%
Remembering that all the Wilson lines are in the adjoint representation and thus real, it is easy to realize that
\beq
\label{adjtdipole_sym}
S_A(-\r,\b)&=&S_A(\r,\b)\; ,\\
\label{operator1_sym}
{\cal O}^j_{(1)}(\r,\b)&=&\frac{1}{N^2_c-1}\left\langle\tr\bigg[\U^{j \dagger}_{(1)}\left(L^+,0;\b+\frac{\r}{2}\right)\, \U\left(L^+,0;\b-\frac{\r}{2}\right)\bigg] \right\rangle_{A}\; ,\\
\label{operator2_sym}
{\cal O}_{(2)}(\r,\b)&=&\frac{1}{N^2_c-1}\left\langle\tr\bigg[\U^{\dagger}_{(2)}\left(L^+,0;\b+\frac{\r}{2}\right)\, \U\left(L^+,0;\b-\frac{\r}{2}\right)\bigg] \right\rangle_{A}\, .
\eeq
Using these properties, the square of the reduced amplitude can be written as
\beq
\hspace{-0.8cm}
&&\frac{1}{N_c^2\!-\!1}
\sum_{\lambda\, \textrm{phys.}} \left\langle  \overline{{\cal M}}^{a b}_{\lambda}(\underline{k}, \q)^{\dag}\;   \overline{{\cal M}}^{a b}_{\lambda}(\underline{k}, \q)  \right\rangle_{A} = \frac{1}{\k^2\,\q^2}
\int d^2\b\; \int d^2\r \; e^{-i (\k-\q)\cdot \r}\;\nonumber\\
&& \times\Bigg\{4\,(\k\!-\!\q)^2
S_A(\r,\b)+2 \frac{L^+}{k^+}\Big[ (\k\!-\!\q)^2\, \k^j + \k^2 (\k^j\!-\!\q^j)\Big]  \bigg[ {\cal O}^j_{(1)}(\r,\b)+  {\cal O}^j_{(1)}(-\r,\b)  \bigg]\nonumber\\
&&\qquad\qquad +2i\frac{L^+}{k^+}\k\cdot(\k-\q)\bigg[ {\cal O}_{(2)}(\r,\b)-  {\cal O}_{(2)}(-\r,\b)  \bigg]     + O\left(\left(\frac{L^+}{k^+}\, \partial^2_{\perp}\right)^2 \right)
 \Bigg\}\; .
\label{ampl_red_squared_eik}
\eeq

Performing the averaging $\langle \dots \rangle_{A}$ over the target restores the invariance under rotations in the transverse plane around the center of the nucleus, so that the quantities $S_A(\r,\b)$, ${\cal O}^j_{(1)}(\r,\b)$ and ${\cal O}_{(2)}(\r,\b)$ should be covariant under such rotations. In particular, ${\cal O}_{(2)}(\r,\b)$ is a scalar with respect to such rotations, and thus it should satisfy the property
\beq
  \int d^2\b\;  {\cal O}_{(2)}(\r,\b) = f_2 (\r^2) \label{O_2_rot}
\eeq
with $f_2$ being an arbitrary function. Indeed, a scalar quantity 
which is a function of a single transverse vector can depend only on 
its modulus due to rotational invariance. Therefore, the two ${\cal O}_{(2)}$ terms in Eq.\eqref{ampl_red_squared_eik} are canceling each other identically upon integration over $\b$.
By contrast, ${\cal O}^j_{(1)}(\r,\b)$ is a vector quantity under these rotations, and thus has to satisfy the property
\beq
 \int d^2\b\;  {\cal O}^j_{(1)}(\r,\b) =   f_1(\r^2)\;   \r^j  \label{O_1_rot}
\eeq
 with $f_1$ being an arbitrary function. Hence, one gets
\beq
 \int d^2\b\;   \bigg[ {\cal O}^j_{(1)}(\r,\b)+  {\cal O}^j_{(1)}(-\r,\b)  \bigg] = 0\, ,
\eeq
 so that the two ${\cal O}^j_{(1)}$ terms in Eq.\eqref{ampl_red_squared_eik} cancel as well.

All in all, the single inclusive gluon cross section writes
\beq
k^+ \frac{d\sigma}{dk^+\, d^2\k} &=&\frac{1}{ \k^2}\!\int \frac{d^2 \q}{(2\pi)^2}\,
 \varphi_p(\q;x_{\textrm{cut}})\;  (\k\!-\!\q)^2  \int d^2\b\! \int d^2\r\,  e^{-i (\k-\q)\cdot \r}\   S_A(\r,\b)\nonumber\\
 &&\hspace{3cm}  +\: O\left(\left(\frac{L^+}{k^+}\, \partial^2_{\perp}\right)^2 \right)
.
\label{cross-section_eik_exp}
\eeq
This is the well-known $\k_\perp$-factorization formula for the single inclusive gluon cross section. Therefore, we have shown that the corrections beyond the eikonal approximation related to the finite size of the target vanish at first order for this particular observable. Higher order terms in the eikonal expansion come with additional transverse derivatives of the background field, denoted by $\partial_{\perp}$ in the expansion parameter. Qualitatively, each $\partial_{\perp}$ can be interpreted as a power of the saturation scale $Q_s$, after averaging over the target. The first non-vanishing corrections, suppressed by $\left(L^+ Q_s^2/k^+\right)^2$, can be calculated systematically by pushing the expansion of the background propagator $\G^{ab}_{k^+}(\underline{x};\underline{y})$ performed in Sec. \ref{sec:next2eik_propag} to higher orders, but we leave this task for future studies.

\subsection{Single transverse spin asymmetry: polarized target}
\label{sec: STSA}
Next-to-eikonal corrections to single inclusive gluon cross section vanish  due to the fact that the quantities in Eqs. \eqref{O_2_rot} and \eqref{O_1_rot}
depend only on one transverse vector. In order to get a non-vanishing contribution at the order $L^+/k^+$, we can consider the single transverse spin asymmetry (SSA) for the process $p+A^{\uparrow}\rightarrow g+X$ i.e. with a transversely polarized target. The SSA is defined as
\beq
\label{SSA_def}
 A_N =  \frac{k^+ \frac{d\sigma^{\uparrow}}{dk^+\, d^2\k} - k^+ \frac{d\sigma^{\downarrow}}{dk^+\, d^2\k} }{ k^+\frac{d\sigma^{\uparrow}}{dk^+\, d^2\k} +  k^+\frac{d\sigma^{\downarrow}}{dk^+\, d^2\k}  }   =   \frac{k^+ \frac{d\sigma^{\uparrow}}{dk^+\, d^2\k} - k^+ \frac{d\sigma^{\downarrow}}{dk^+\, d^2\k} }{2  k^+\frac{d\sigma}{dk^+\, d^2\k}  }\; ,
\eeq
where the denominator corresponds to the unpolarized cross section, Eq.\eqref{cross-section_eik_exp}.

In the CGC, it is not  yet known how to include the transverse polarization of the target in practice\footnote{However, recent progresses have been performed in that direction in the context of the Glauber-Mueller approximation \cite{Kovchegov:2013cva}.}. However, it is clear that the dependence on the transverse spin vector $\s$ should appear through the probability distribution for the background field, leading to $\s$-dependent target-averaged quantities such as $S_A$, $ {\cal O}^j_{(1)}$ and ${\cal O}_{(2)}$. Then, due to transverse rotational symmetry around the center of the target, one gets
\beq
\label{adjtdipole_rot_s}
S_A(-\r,-\b,-\s)&=&S_A(\r,\b,\s),\\
\label{operator1_rot_s}
{\cal O}^j_{(1)}(-\r,-\b,-\s) &=&- {\cal O}^j_{(1)}(\r,\b,\s),\\
\label{operator2_rot_s}
{\cal O}_{(2)}(-\r,-\b,-\s)&=& {\cal O}_{(2)}(\r,\b,\s)\; .
\eeq
Eq.\eqref{adjtdipole_sym} is actually valid for each background field, before taking the target average. Hence, it generalizes as
\beq
\label{adjtdipole_s}
S_A(-\r,\b,\s)&=&S_A(\r,\b,\s)
\eeq
in the case of a transversely polarized target. Eqs.\eqref{cross-section_3} and \eqref{ampl_red_squared_eik} are still valid in the transversely polarized target case, after including the $\s$ dependence in $S_A$, $ {\cal O}^j_{(1)}$ and ${\cal O}_{(2)}$. The strict eikonal contribution to the numerator of the SSA can be evaluated by using Eq.\eqref{adjtdipole_rot_s}, the change of variable $\b\mapsto -\b$ in the second term and finally Eq.\eqref{adjtdipole_s}, as follows
\beq
 \int d^2\b \Big[  S_A(\r,\b,\s)-S_A(\r,\b,-\s)\Big] &=& \int d^2\b \Big[  S_A(\r,\b,\s)-S_A(-\r,-\b,\s)\Big] \nonumber\\
 &=& \int d^2\b \Big[  S_A(\r,\b,\s)-S_A(-\r,\b,\s)\Big]\nonumber\\
 &=& 0\; .
\eeq
In the same way, one can simplify the next-to-eikonal contributions to the numerator of the SSA as
\beq
 && \int d^2\b \Big[{\cal O}^j_{(1)}(\r,\b,\s)+{\cal O}^j_{(1)}(-\r,\b,\s)-{\cal O}^j_{(1)}(\r,\b,-\s)-{\cal O}^j_{(1)}(-\r,\b,-\s)\Big]\nonumber\\
  &=& 2 \int d^2\b \Big[ {\cal O}^j_{(1)}(\r,\b,\s)+{\cal O}^j_{(1)}(-\r,\b,\s)\Big]
\eeq
and
\beq
 && \int d^2\b \Big[{\cal O}_{(2)}(\r,\b,\s)-{\cal O}_{(2)}(-\r,\b,\s)-{\cal O}_{(2)}(\r,\b,-\s)+{\cal O}_{(2)}(-\r,\b,-\s)\Big]\nonumber\\
  &=& 2 \int d^2\b \Big[ {\cal O}_{(2)}(\r,\b,\s)-{\cal O}_{(2)}(-\r,\b,\s)\Big]\; .
\eeq
Therefore, the expression for the numerator of SSA reads
\beq
\label{SSA_num}
&&\hspace{-1.5cm}
k^+\left(  \frac{d\sigma^{\uparrow}}{dk^+\, d^2\k} -  \frac{d\sigma^{\downarrow}}{dk^+\, d^2\k} \right)=\frac{2}{\k^2}\frac{L^+}{k^+}\!\int \frac{d^2 \q}{(2\pi)^2}\
 \varphi_p(\q;x_{\textrm{cut}})  \nonumber\\
 &&\times
 \Bigg\{\Big[ (\k\!-\!\q)^2\, \k^j + \k^2 (\k^j\!-\!\q^j)\Big] \int d^2\r \, \cos \Big(\r\cdot(\k\!-\!\q)\Big)\int d^2\b \;  {\cal O}^j_{(1)}(\r,\b,\s)\nonumber\\
 &&
 +\k\cdot(\k\!-\!\q)  \int d^2\r \, \sin \Big(\r\cdot(\k\!-\!\q)\Big)\int d^2\b \;  {\cal O}_{(2)}(\r,\b,\s) \Bigg\} + O\left(\left(\frac{L^+}{k^+}\, \partial^2_{\perp}\right)^2 \right)\, .
\eeq
Note that in the case of SSA, the strict eikonal term vanishes whereas the next-to-eikonal contributions are in general nonzero. This is the reversed situation compared to the un-polarized cross section, Eq.\eqref{cross-section_eik_exp}. This feature of the eikonal expansion is reminiscent of the behaviour of the twist expansion: unpolarized cross sections typically receive twist 2 but no twist 3 contributions, whereas in SSAs twist 2 contributions vanish, leaving the twist 3 as the leading contributions \cite{Efremov:1984ip,Qiu:1991pp,Qiu:1998ia}.

\subsection{Cross section for polarized gluon production}
\label{sec: polarized gluon}

In practice, it might be challenging to produce a beam of transversely polarized large nuclei. However, similar physics can be probed by studying the production of polarized hadrons in unpolarized pA collisions.

There are typically two mechanisms leading to polarized hadron production in unpolarized pA collisions. One possibility is that a polarized quark or gluon is produced, which then fragments in a standard way to a polarized hadron. The other possibility is that the polarization is induced during the fragmentation process. That second possibility has been studied in the CGC framework in the case of transversely polarized hyperon production \cite{Boer:2002ij}. By contrast, the focus on the present section is on the first type of mechanism.
For simplicity, let us restrict ourselves to the study of the asymmetry in the light-front helicity of the produced gluon, assuming that this asymmetry is preserved at the hadron level by fragmentation.

The calculation of that asymmetry is almost identical to the calculation on the single inclusive gluon cross section, Eq.\eqref{cross-section_3}, except that instead of summing over the helicity $\lambda=\pm 1$ of the produced gluon,
one takes the difference between the $\lambda=+1$ and $\lambda=-1$ contributions, i.e.
\beq
& & \hspace{-1cm} k^+\, \frac{d\sigma^+}{dk^+\, d^2\k}-k^+\, \frac{d\sigma^-}{dk^+\, d^2\k}=\int \frac{d^2 \q}{(2\pi)^2}\;
 \varphi_p(\q;x_{\textrm{cut}})\;  \nonumber\\
 && \hspace{4cm} \times\: \frac{ \q^2}{4}\;
\frac{1}{N_c^2\!-\!1}
\sum_{\lambda\, \textrm{phys.}} \lambda\; \left\langle  \overline{{\cal M}}^{a b}_{\lambda}(\underline{k}, \q)^{\dag}\;   \overline{{\cal M}}^{a b}_{\lambda}(\underline{k}, \q)  \right\rangle_{A} .
\label{gluon_asym_cross_section}
\eeq
The difference between the two light-front helicity states can be calculated thanks to the identity
\beq
\sum_{\lambda\, \textrm{phys.}} \lambda\;  \varepsilon^{i}_\lambda\; \varepsilon^{*j}_\lambda = -i\; \epsilon^{ij}\, ,
\label{helicity_difference}
\eeq
where $\epsilon^{ij}$ is the antisymmetric matrix with $\epsilon^{12}=+1$.

Inserting the eikonal expansion of the reduced amplitude, Eq.\eqref{totredexp}, into Eq.\eqref{gluon_asym_cross_section}, one obtains
\beq
& & \hspace{-1cm} k^+\, \frac{d\sigma^+}{dk^+\, d^2\k}-k^+\, \frac{d\sigma^-}{dk^+\, d^2\k}=\frac{L^+}{k^+}\int \frac{d^2 \q}{(2\pi)^2}\;
 \varphi_p(\q;x_{\textrm{cut}})\; \q^2  \int d^2\b\; \int d^2\r \; e^{-i (\k-\q)\cdot \r}\;\nonumber\\
&& \times\Bigg\{- i\left[ \left(\frac{\k^i}{\k^2}\!-\!\frac{\q^i}{\q^2}\right)\, \epsilon^{ij}
-2\:  \frac{\left(\epsilon^{il}\, \k^i\, \q^l\right)}{\k^2\, \q^2} \, \k^j\right]
{\cal O}^j_{(1)}(\r,\b)\nonumber\\
&& \hspace{3cm}\qquad\qquad -\,  \frac{(\epsilon^{ij}\, \k^i\, \q^j)}{\k^2\, \q^2}\;
 {\cal O}_{(2)}(\r,\b)
 \Bigg\}    + O\left(\left(\frac{L^+}{k^+}\, \partial^2_{\perp}\right)^2 \right)\; .
\label{gluon_asym_cross_section_result}
\eeq
In order to arrive at Eq.\eqref{gluon_asym_cross_section_result}, we made use of the symmetry properties of ${\cal O}^j_{(1)}(\r,\b)$ and ${\cal O}_{(2)}(\r,\b)$ discussed in Sec. \ref{sec: next2eik_sigma}. As in the case of the SSA discussed in the previous section, the term at eikonal order vanishes in Eq.\eqref{gluon_asym_cross_section_result} and the next-to-eikonal contributions dominate. This shows that the next-to-eikonal operators ${\cal O}^j_{(1)}(\r,\b)$ and ${\cal O}_{(2)}(\r,\b)$ are important not only for transverse spin but also for longitudinal spin  physics.
Moreover, ${\cal O}^j_{(1)}(\r,\b)$ and ${\cal O}_{(2)}(\r,\b)$ can be probed in unpolarized pA collisions, by studying polarized hadron production. The case of a polarized gluon considered in this section is a proxy for longitudinally polarized hadron production. However, one can expect that the same operators ${\cal O}^j_{(1)}(\r,\b)$ and ${\cal O}_{(2)}(\r,\b)$ should give a dominant contribution also in the case of transversely polarized  hadron production.

\subsection{Eikonal expansion of the single inclusive gluon spectrum}
\label{sec: analysis}
Alternatively, one can
consider the single inclusive gluon spectrum whose general expression is given in Eq.\eqref{dN_mom_2}, for a given impact parameter $\B$. One should average over the azimuthal direction of the impact parameter $\B$, since it is not possible\footnote{In heavy ion and pA collisions the events are sorted by the experimental collaborations according to measurements of particle multiplicity or deposited energy at forward rapidities. In the case of heavy ion collisions, there is a strong correlation between the impact parameter and such forward observables, which makes the magnitude and also the azimuthal direction of $\B$ accessible experimentally. Thus, events in the same class (called centrality class) have similar $|\B|$. In the case of pA collisions, the yields and the azimuthal correlations \cite{Aad:2013fja,Chatrchyan:2013nka,ABELEV:2013wsa} of produced particles are expected to be driven by event-by-event fluctuations and not by the average geometry of the collision.
Hence, the experimental reconstruction of the azimuthal direction of $\B$ seems hopeless in pA collisions. Moreover, the binning 
of events according to the forward activity 
is not such a good approximation of the binning of events according to the magnitude $|\B|$ of their impact parameter. However, let us consider in this section pA collisions with a given $|\B|$. } to measure it experimentally, which leads to
\beq
&& k^+\, \frac{dN}{dk^+\, d^2\k} \big(|\B|\big)=\int \frac{d^2 \q}{(2\pi)^2}\; \int \frac{d^2 \bf{\Delta}}{(2\pi)^2}\; \text{J}_0\Big(|\B|\,|\Delta|\Big)\;
\widetilde{{\cal W}}(\q,\Delta;x_{\textrm{cut}})\;
\nonumber\\
&& \hspace{2cm} \times  \frac{\q^2}{4}\, \frac{1}{N_c^2\!-\!1}\, \sum_{\lambda\, \textrm{phys.}} \left\langle  \overline{{\cal M}}^{a b}_{\lambda}\left(\underline{k}, \q\!-\!\frac{\bf{\Delta}}{2} \right)^{\dag}\;   \overline{{\cal M}}^{a b}_{\lambda}\left(\underline{k}, \q\!+\!\frac{\bf{\Delta}}{2}\right)  \right\rangle_{A}\, ,
\label{dN_mom_3}
\eeq

where
\beq
\widetilde{{\cal W}}(\q,\Delta;x_{\textrm{cut}})\, =  \int d^2 \b_p  \; e^{-i   \bf{\Delta}\cdot\b_p}\;
{\cal W}(\q,\b_p;x_{\textrm{cut}}) \; .
\eeq
Using the expression for the total gluon-nucleus reduced amplitude, $\overline{{\cal M}}^{a b}_{\lambda}(\underline{k}, \q)$, Eq.\eqref{totredexp}, and the notations introduced in Eqs.\eqref{adjtdipole}, \eqref{operator1} and \eqref{operator2}, one obtains
\beq
\label{shiftedM2}
&&
\frac{1}{N_c^2\!-\!1}\, \sum_{\lambda\, \textrm{phys.}} \left\langle  \overline{{\cal M}}^{a b}_{\lambda}\left(\underline{k}, \q\!-\!\frac{\bf{\Delta}}{2} \right)^{\dag}\;   \overline{{\cal M}}^{a b}_{\lambda}\left(\underline{k}, \q\!+\!\frac{\bf{\Delta}}{2}\right)  \right\rangle_{A}  =
\int d^2\r \, e^{-i(\k-\q)\cdot\r} \int d^2\b \, e^{i\Delta\cdot\b} \nonumber\\
&&\times
\Bigg\{
4M_0^i(\k,\q,\Delta) \; M_0^i(\k,\q,-\Delta) \; S_A(\r,\b) \nonumber\\
&&
+ \frac{2L^+}{k^+} \, \Big[  M_0^i(\k,\q,-\Delta) \, M_1^{ij}(\k,\q,\Delta) \, {\cal O}^j_{(1)}(\r,\b) + M_0^i(\k,\q,\Delta) \, M_1^{ij}(\k,\q,-\Delta) \, {\cal O}^j_{(1)}(-\r,\b) \Big] \nonumber\\
&&
-i\frac{2L^+}{k^+} \,  \Big[  M_0^i(\k,\q,-\Delta) \, M_2^{i}(\q,\Delta) \, {\cal O}_{(2)}(\r,\b) - M_0^i(\k,\q,\Delta) \, M_2^{i}(\q,-\Delta) \, {\cal O}_{(2)}(-\r,\b) \Big] \nonumber\\
&&\hspace{7cm}
+ \; O\left(\left(\frac{L^+}{k^+}\, \partial^2_{\perp}\right)^2 \right)
\Bigg\} \; ,
\eeq
with the following notations:
\beq
\label{M0i}
M_0^i(\k,\q,\Delta) & = &\frac{\k^i}{\k^2}-\frac{\q^i+\frac{\Delta^i}{2}}{\left(\q+\frac{\Delta}{2}\right)^2}\ ,\\
\label{M1ij}
M_1^{ij}(\k,\q,\Delta) & = & \delta^{ij} - 2 \, \frac{ \left( \q^i+\frac{\Delta^i}{2} \right) }{\left(\q+\frac{\Delta}{2}\right)^2} \; \k^j\ , \\
\label{M2i}
M_2^{i}(\q,\Delta) & = & \frac{\q^i+\frac{\Delta^i}{2}}{\left(\q+\frac{\Delta}{2}\right)^2} \; .
\eeq

As discussed in Sec. \ref{sec: next2eik_sigma}, the quantities $S_A(\r,\b)$, ${\cal O}^j_{(1)}(\r,\b)$ and ${\cal O}_{(2)}(\r,\b)$ are covariant under rotations in the transverse plane after averaging over the target configuration. $S_A(\r,\b)$ and ${\cal O}_{(2)}(\r,\b)$ are scalars and ${\cal O}^j_{(1)}(\r,\b)$ is a vector under such rotations, so that
\beq
\label{adjtdipole_rot}
S_A(-\r,-\b)&=&S_A(\r,\b)\; ,\\
\label{operator1_rot}
{\cal O}^j_{(1)}(-\r,\b) &=&- {\cal O}^j_{(1)}(\r,-\b)\; ,\\
\label{operator2_rot}
{\cal O}_{(2)}(-\r,\b) &=& {\cal O}_{(2)}(\r,-\b)\; .
\eeq
Combining Eqs.\eqref{adjtdipole_sym} and \eqref{adjtdipole_rot},  one also obtains
\beq
\label{adjtdipole_flip_b}
S_A(\r,-\b)&=&S_A(\r,\b)\; .
\eeq
Thanks to Eqs.\eqref{operator1_rot} and \eqref{operator2_rot}, Eq.\eqref{shiftedM2}  can now be written as
\beq
\label{shiftedM2_Mom}
&&
\hspace{-2cm}\frac{1}{N_c^2\!-\!1}\, \sum_{\lambda\, \textrm{phys.}} \left\langle  \overline{{\cal M}}^{a b}_{\lambda}\left(\underline{k}, \q\!-\!\frac{\bf{\Delta}}{2} \right)^{\dag}\;   \overline{{\cal M}}^{a b}_{\lambda}\left(\underline{k}, \q\!+\!\frac{\bf{\Delta}}{2}\right)  \right\rangle_{A}  =\nonumber\\
&&\hspace{-1cm}\times
\Bigg\{
4M_0^i(\k,\q,\Delta) \; M_0^i(\k,\q,-\Delta) \; \widehat{S_A}(\k\!-\!\q,\Delta) \nonumber\\
&&
+ \frac{2L^+}{k^+} \, \Big[  M_0^i(\k,\q,-\Delta) \, M_1^{ij}(\k,\q,\Delta) \, \widehat{{\cal O}^j_{(1)}}(\k\!-\!\q,\Delta) \nonumber\\
&&
\hspace{2cm}- M_0^i(\k,\q,\Delta) \, M_1^{ij}(\k,\q,-\Delta) \, \widehat{{\cal O}^j_{(1)}}(\k\!-\!\q,-\Delta)  \Big] \nonumber\\
&&
-i\frac{2L^+}{k^+} \,  \Big[  M_0^i(\k,\q,-\Delta) \, M_2^{i}(\q,\Delta) \, \widehat{{\cal O}_{(2)}}(\k\!-\!\q,\Delta) \nonumber\\
&&
\hspace{2cm} - M_0^i(\k,\q,\Delta) \, M_2^{i}(\q,-\Delta) \, \widehat{{\cal O}_{(2)}}(\k\!-\!\q,-\Delta) \Big] \nonumber\\
&&\hspace{5cm}
+ \; O\left(\left(\frac{L^+}{k^+}\, \partial^2_{\perp}\right)^2 \right)
\Bigg\} \; ,
\eeq
with the notation
\beq
\label{Full_FT}
\widehat{{\cal F}}(\k\!-\!\q,\Delta) = \int d^2\r \, e^{-i(\k-\q)\cdot\r} \int d^2\b \, e^{i\Delta\cdot\b}\;  {\cal F}(\r,\b)
\eeq
for the Fourier transform to full momentum space. Eq.\eqref{adjtdipole_flip_b} shows that $\widehat{S_A}(\k\!-\!\q,\Delta) $ is even in $\Delta$, and thus the strict eikonal term in Eq.\eqref{shiftedM2_Mom}  is even in $\Delta$. On the other hand, the next-to-eikonal corrections in Eq.\eqref{shiftedM2_Mom} are obviously odd in $\Delta$. When inserting Eq.\eqref{shiftedM2_Mom}  into the single inclusive gluon spectrum, Eq.\eqref{dN_mom_3}, the eikonal term involves only the even piece of $\widetilde{{\cal W}}(\q,\Delta;x_{\textrm{cut}})$ with respect to $\Delta$, whereas next-to-eikonal corrections involve only the odd piece of $\widetilde{{\cal W}}(\q,\Delta;x_{\textrm{cut}})$.

Note that the existence of an odd piece of $\widetilde{{\cal W}}(\q,\Delta;x_{\textrm{cut}})$ depends on the employed approximations. For example, the large nucleus approximation, $\B+\b_p\simeq \B$, leading to Eq.\eqref{dN_mom_large_A}, amounts to make the replacement
\beq
\widetilde{{\cal W}}(\q,\Delta;x_{\textrm{cut}})\to \widetilde{{\cal W}}(\q,\Delta=0;x_{\textrm{cut}})=\varphi_p(\q;x_{\textrm{cut}})
\eeq
at the level of Eq.\eqref{dN_mom_3}.
Therefore, this approximation neglects the odd piece of $\widetilde{{\cal W}}(\q,\Delta;x_{\textrm{cut}})$ with respect to $\Delta$. As a result, next-to-eikonal corrections vanish within the large nucleus approximation.

Within the CGC formalism, the Wigner distribution of Weizs\"acker-Williams gluons in the projectile, ${\cal W}(\q,\b_p;x_{\textrm{cut}})$, is defined via Eq.\eqref{rhorho_vs_Wigner}. The ordering of color charge densities is irrelevant in a two point correlator, which ensures that the correlator in Eq.\eqref{rhorho_vs_Wigner} is invariant under $\r_p\mapsto -\r_p$. As a result, the Wigner distribution, ${\cal W}(\q,\b_p;x_{\textrm{cut}})$, is invariant under $\q\mapsto -\q$ when calculated within the CGC. On the other hand, thanks to the invariance under rotations in the transverse plane around the center of the projectile, one has ${\cal W}(\q,\b_p;x_{\textrm{cut}})={\cal W}(-\q,-\b_p;x_{\textrm{cut}})$. These two invariances imply that  ${\cal W}(\q,\b_p;x_{\textrm{cut}})$ is even in $\b_p$, and thus that $\widetilde{{\cal W}}(\q,\Delta;x_{\textrm{cut}})$ is even in $\Delta$.

There are different ways of generating an odd piece in the description of the projectile.
First, one can refine the description of the projectile by going beyond the eikonal accuracy, or equivalently beyond the Weizs\"acker-Williams approximation for the gluons involved in the process. Then, one would have the full Wigner distribution of gluons in the projectile, which has an odd piece in $\Delta$ \cite{Lorce:2013pza}.

Second, within the Weizs\"acker-Williams regime in the CGC formalism, the projectile can couple to the rest of the process via more than two gluons.
At the three-gluon exchange level, the projectile can couple in an odd way, via the Odderon\footnote{The Odderon should  lead to non-vanishing next-to-eikonal corrections to unpolarized observables for the same reason as it leads to non-vanishing SSA at eikonal accuracy \cite{Kovchegov:2012ga}.}. However, the coupling of three gluons to the projectile can arise only at next-to-leading order in the coupling $g$, and beyond. The investigation of that possibility is therefore beyond the scope of the present paper.


Third, one can consider the case of a transversely polarized projectile, \emph{i.e.} the process $p^{\uparrow}+A\rightarrow g+X$, and study the following observable
\beq
&& k^+\, \frac{dN^{\uparrow}}{dk^+\, d^2\k} \big(|\B|\big)-k^+\, \frac{dN^{\downarrow}}{dk^+\, d^2\k} \big(|\B|\big)=\int \frac{d^2 \q}{(2\pi)^2}\; \int \frac{d^2 \bf{\Delta}}{(2\pi)^2}\; \text{J}_0\Big(|\B|\,|\Delta|\Big)\;\nonumber\\
&& \hspace{4cm}  \times \bigg[\widetilde{{\cal W}}(\q,\Delta,\s ;x_{\textrm{cut}})-\widetilde{{\cal W}}(\q,\Delta,-\s ;x_{\textrm{cut}})\bigg]\;
\nonumber\\
&& \hspace{2cm} \times  \frac{\q^2}{4}\, \frac{1}{N_c^2\!-\!1}\, \sum_{\lambda\, \textrm{phys.}} \left\langle  \overline{{\cal M}}^{a b}_{\lambda}\left(\underline{k}, \q\!-\!\frac{\bf{\Delta}}{2} \right)^{\dag}\;   \overline{{\cal M}}^{a b}_{\lambda}\left(\underline{k}, \q\!+\!\frac{\bf{\Delta}}{2}\right)  \right\rangle_{A}
\label{dN_mom_polarized_proj}\; .
\eeq
The relation $\widetilde{{\cal W}}(\q,\Delta,-\s ;x_{\textrm{cut}})=\widetilde{{\cal W}}(-\q,-\Delta,\s ;x_{\textrm{cut}})$ holds thanks to the invariance under transverse rotations. The invariance under permutation of the two color charge densities should still be valid in the polarized projectile case, leading to $\widetilde{{\cal W}}(-\q,\Delta,\s ;x_{\textrm{cut}})=\widetilde{{\cal W}}(\q,\Delta,\s ;x_{\textrm{cut}})$. Using these two relations, it is clear that only the piece of $\widetilde{{\cal W}}(\q,\Delta,\s ;x_{\textrm{cut}})$ which is odd in $\Delta$ contributes to the SSA observable, Eq.\eqref{dN_mom_polarized_proj}. Therefore, this SSA is also dominated by next-to-eikonal contributions involving the operators ${\cal O}^j_{(1)}(\r,\b)$ and ${\cal O}_{(2)}(\r,\b)$.

\section{Conclusions}
\label{sec:conclusions}

In this paper we have performed, for the first time, a systematic eikonal expansion of the retarded gluon propagator in a background field which is one of the most crucial building blocks of the high energy dense-dilute scattering processes, and also of medium-induced gluon radiation. The leading term in the expansion - corresponding to the strict eikonal limit - is given by a Wilson line operator, as it is well-known. However, the next-to-eikonal terms in the expansion involve new operators which include gradients of the background field, referred to as \emph{decorated Wilson lines}. The explicit forms of these new operators  in the light-cone gauge are given in Eqs. \eqref{decoratedWilson1A} and \eqref{decoratedWilson2A}.

The expansion is then applied to single inclusive gluon production in pA collisions in order to study the corrections to the CGC beyond the eikonal limit, focusing on finite target size effects for dense-dilute scattering processes. The first observable considered is the single inclusive gluon cross section. The strict eikonal term provides the usual $k_{\perp}$-factorization formula, whereas the first next-to-eikonal corrections vanish for this particular observable, extending the validity range of the $k_{\perp}$-factorization formula.

Various types of spin asymmetries for gluon production in pA collisions, such as single transverse spin asymmetry for a polarized target, or helicity asymmetry for the produced gluon in unpolarized pA collisions, have been also considered. In each of these, the eikonal contribution vanishes, leaving the next-to-eikonal correction as the leading term. 
This result shows the analogy between the twist-3 contributions to hard processes and next-to-eikonal contributions to high-energy processes.

The eikonal expansion of the retarded gluon propagator is also applied to single inclusive gluon spectrum for pA collisions at fixed impact parameter. This observable is usually considered in the literature within a large nucleus approximation. It is shown that for the single inclusive gluon spectrum within the large nucleus approximation, the next-to-eikonal contributions vanish. However, without adopting the large nucleus approximation, the next-to eikonal corrections
exhibit a different behaviour. Upon integration over the azimuthal direction of the impact parameter $\B$, the leading eikonal contribution to the single inclusive gluon spectrum involves the even piece of the Wigner distribution of gluons inside the projectile with respect to the momentum transfer $\Delta$. The next-to-eikonal contribution, however, involves the odd piece of this Wigner distribution with respect to $\Delta$.

The first non-vanishing contribution to the strict eikonal limit of the single inclusive gluon cross section and the spectrum with large nucleus approximation arises at the order of $\left(L^+ Q_s^2/k^+\right)^2$ which can be obtained systematically by continuing the expansion performed in this paper to higher orders. This issue is left for future studies.

The eikonal expansion introduced in this work has further possible applications. On the one hand, it determines next-to-eikonal corrections to other dense-dilute high-energy scattering processes like DIS and single inclusive gluon production in the hybrid formalism. On the other hand, it can provide new insight into jet quenching physics, since the medium induced gluon radiation occurs only beyond the strict eikonal limit.

Given the fact that the decorated Wilson lines follow a light-like trajectory, one can expect the corresponding decorated dipole operators to have rapidity divergences. This naturally raises the question of the low $x$ evolution of the new operators, which requires dynamics beyond the JIMWLK evolution. This crucial issue is also left for future studies.

So far, we have studied only one type of corrections with respect to the eikonal limit, namely the ones associated with the finite length of the target. However, there are other types of power-suppressed corrections at high-energy, such as the ones associated with the $x^-$ dependence of the background field representing the target or with the transverse components of this background field. Such corrections have to be studied in order to build the complete eikonal expansion at high energy in a systematic way. 

\section*{Acknowledgements}
We thank Alexis Moscoso for pointing out a mistake in the calculation, Bernard Pire for useful suggestions, and C\'edric Lorc\'e for helpful correspondence.
We also thank M. Braun, G. Chirilli, A. Dumitru, Y. Kovchegov, A. Kovner and G. Milhano for discussions. This work is supported by European Research Council grant HotLHC ERC-2011-StG-279579; by People Programme (Marie Curie Actions) of the European Union Seventh Framework Programme FP7/2007-2013/ under REA grant agreement no. 318921; by Ministerio de Ciencia e Innovaci\'on of Spain under projects FPA2009-11951 and FPA2011-22776, and the Consolider-Ingenio 2010 Programme CPAN (CSD2007-00042);  by Xunta de Galicia (GRC2013-024); and by FEDER. TA and MM thank to the CERN Theory division for their kind hospitality where part of this project was done.  MM is supported by U.S. Department of Energy under Grant No. DE-SC0004286. MM thanks to the physics department at the Universidade de Santiago de Compostela for the financial support at the early stages of this work.

\appendix

\bibliography{biblio}{}
\bibliographystyle{jhep}

\end{document}